\providecommand{\reviewmode}{0}
\ifnum\reviewmode=1
  \documentclass[preprint,12pt]{elsarticle}
\else
  \documentclass[final,5p,times,twocolumn]{elsarticle}
\fi

\ifnum\reviewmode=1
  \usepackage{lineno}             %
  \AtBeginDocument{\linenumbers}
  \usepackage{setspace}           %
\fi
\usepackage{adjustbox}            %
\usepackage[T1]{fontenc}
\usepackage[utf8]{inputenc}      %
\usepackage{amsmath, amssymb}
\usepackage{graphicx}
\usepackage{booktabs}
\usepackage[separate-uncertainty=true, detect-weight=true]{siunitx}
\usepackage[version=4]{mhchem}
\usepackage{microtype}
\usepackage{enumitem}            %
\usepackage[hidelinks]{hyperref}

\hypersetup{
  pdftitle  = {Predicting the experimental lattice thermal conductivity of half-Heusler
               compounds by machine learning and transfer-function calibration},
  pdfauthor = {Shivam Randive, Ojaswini Randive, Heet Lungaria, Himanshu Pandey},
  pdfsubject= {Lattice thermal conductivity of half-Heusler thermoelectrics},
  pdfkeywords = {half-Heusler, lattice thermal conductivity, thermoelectric,
                 Boltzmann transport equation, machine learning, transfer-function calibration,
                 experimental validation}}

\newcommand{\pandeyemail}{hp@phy.svnit.ac.in}
\newcommand{\repourl}{\href{https://doi.org/10.5281/zenodo.24335391}%
  {\texttt{doi:10.5281/zenodo.24335391}} (concept DOI
  \href{https://doi.org/10.5281/zenodo.22642665}{\texttt{10.5281/zenodo.22642665}}, which resolves
  to the latest version)}
\newcommand{\acknowledgementtext}{
  Google Cloud credits funded the large-language-model extraction of measurements from primary
  publications (Section~\ref{sec:dataset}). We thank the maintainers of the databases this work
  rests on (Starrydata, AFLOW, the Materials Project, JARVIS, OQMD, DxMag, the Crystallography Open
  Database and the OPTIMADE consortium), whose curation made an experiment-anchored corpus of this
  kind possible. We also thank the authors of the \num{189} experimental and \num{29} computational
  studies whose measurements and calculations make up the corpus analysed here.}

\journal{Computational Materials Science}

\begin{document}

\begin{frontmatter}

\title{Predicting the experimental lattice thermal conductivity of half-Heusler compounds
       by machine learning and transfer-function calibration}

\author[svnit]{Shivam Randive}
\author[nitrr]{Ojaswini Randive}
\author[svnit]{Heet Lungaria}
\author[svnit]{Himanshu Pandey\corref{cor1}}
\ead{\pandeyemail}
\cortext[cor1]{Corresponding author.}

\affiliation[svnit]{organization={Department of Physics, Sardar Vallabhbhai National Institute
                                  of Technology},
                    city={Surat},
                    country={India}}
\affiliation[nitrr]{organization={Department of Metallurgical and Materials Engineering,
                                  National Institute of Technology Raipur},
                    state={Chhattisgarh},
                    country={India}}

\begin{abstract}
High-throughput searches for half-Heusler thermoelectrics increasingly rely on lattice thermal
conductivity ($\kappa_L$) from first-principles calculations, or from machine-learning models trained
on them. These calculations describe an ideal single crystal, whereas measurements are made on
polycrystalline samples, in which grain boundaries and defects scatter heat as well. The resulting gap
between calculated and measured $\kappa_L$ has been noted for individual compounds but not, to our
knowledge, quantified across a materials class, so the values used for screening carry an unknown, systematic bias. Here we
quantify and correct this bias for cubic half-Heusler compounds with 18 valence electrons. Using a curated literature corpus that
keeps calculated and measured values apart, we show that calculations overestimate measured $\kappa_L$
in most bonding families, and that the excess shrinks with temperature. Scattering by defects, added
to the intrinsic scattering between lattice vibrations, would give such a trend, but the data do not
single out one cause. A simple transfer function with this temperature dependence removes the median
offset, and refitting its magnitude within a bonding family lowers the median error further, though not
consistently compound by compound. Combined with a machine-learning model that predicts the calculated
value from crystal structure, and tested on chemistries withheld from all fitting, the method predicts
measured $\kappa_L$ in this class about as closely as independent laboratories agree with one another, and provides testable predictions
for compounds not yet measured.
\end{abstract}


\end{frontmatter}
\ifnum\reviewmode=1 \onehalfspacing\fi

\section{Introduction}
\label{sec:intro}

A thermoelectric generator needs a material that conducts charge freely but obstructs heat. Half
Heuslers meet the first requirement far better than the second. Their electronic performance is
well established, with figures of merit approaching $zT \approx 1.5$ in $p$-type
\ce{NbFeSb}-based~\cite{fu2015realizing} and $zT \approx 1.4$ in \ce{ZrCoBi}-based
alloys~\cite{zhu2018discovery}. Their lattice thermal conductivity $\kappa_L$, however, remains
high, and because $zT$ varies inversely with $\kappa_L$, most advances have come from suppressing
it after synthesis. The screening problem is to identify compounds with intrinsically low
$\kappa_L$ before they are made.

That requires a calculated $\kappa_L$. Solving the phonon Boltzmann
transport equation (BTE) from first principles gives $\kappa_L$ with no experimental input; for
the half Heusler \ce{ZrCoSb} the calculated temperature dependence agrees reasonably with
measurement, running slightly above it, with no fitted parameter~\cite{shiomi2011thermal}. The
obstacle is cost. Carrete \emph{et al.} calculated \num{32} half Heuslers in full and trained a
random-forest model on them to search a far larger set~\cite{carrete2014finding}, and much
subsequent work has approximated the calculation at lower cost, through compressed-sensing
descriptors~\cite{liu2020a} and machine-learning models trained on computed
conductivities~\cite{luo2023predicting,trans2022lattice,miyazaki2021machine,jaafreh2021lattice,bhattacharjee2022thorough}.
Most recently, Paliwal and Alam compared seven tree-based ensemble regressors trained on \num{4127}
calculated values for \num{150} compounds; the best, an extra-trees model, reached
$R^2 = \num{0.9994}$ (on $\log\kappa_L$) in repeated cross-validation on calculated
labels~\cite{paliwal2025accelerate}.

We call such models \emph{surrogates}, because they stand in for the calculation. A surrogate is
accurate against the quantity it was trained on, and that quantity is a calculation, not a
measurement. How far the two differ has been judged in two ways. An earlier tutorial on the method
stated that it ``has enabled thermal conductivity predictions that agree with experimental
measurements for diverse crystalline materials''~\cite{mcgaughey2019phonon}. A later controlled
multi-code comparison, by the same lead author, found the solvers within \SI{15}{\percent} of
their mean values, yet reported that the calculated conductivities ``do not generally show
agreement with experimental measurements, which is attributed to the limitations of the density
functional theory calculations''~\cite{mcgaughey2025phonon}. Paliwal and Alam note that
computed $\kappa_L$ ``often overestimates the experimental values because real samples inevitably
contain impurities, defects, and grain boundaries''~\cite{paliwal2025accelerate}, but they do not
quantify this \emph{offset} between calculation and measurement. What has been missing, to our
knowledge, is the size and sign of the offset across a materials class, and a correction for it
validated against measurements withheld from its fit.

In our corpus, published transport calculations exceed the measured $\kappa_L$ of the same compound
by a median factor of \num{1.69} over \num{34} compounds. A recent large-scale model shows what
that means for a surrogate. The Elemental Spatial Density Neural Network Force Field
(Elemental-SDNNFF)~\cite{rodriguez2023millionsca} replaces first-principles force calculations with
a neural network and so predicts the phonon properties of \num{11866} Heusler structures, a scale
out of reach of direct calculation. For the measured half Heuslers of our validation set that also
have a published first-principles calculation, its values agree closely with that calculation
(median ratio \num{0.99} over \num{18} compounds). Against the laboratory measurements of the
\num{21} it shares with the same set, however, they are a median \num{2.01} times too high
(Section~\ref{sec:sdnnff}). The force field does well what it was built to do; it is the target
itself that lies away from experiment.

The data-driven routes from calculation to measurement that we know of do not correct a full
first-principles calculation.
Zhu \emph{et al.} transfer-learned a crystal-graph network from \num{2668} computed conductivities
onto \num{132} measured values, starting from semi-empirical high-throughput
estimates~\cite{zhu2021charting}. Miller \emph{et al.} fitted a semi-empirical model to
experiment~\cite{miller2017capturing}, Chen \emph{et al.} trained a Gaussian process on about one
hundred measured solids~\cite{chen2019machine}, and a related study pretrained on empirical-model
estimates and then fine-tuned on a single high-fidelity target that pools measurements with
first-principles calculations~\cite{liu2022leveraging}. Pooling hides the offset. For band gaps,
learning the difference between calculation and measurement has been shown to work better than
pooling mixed sources~\cite{hutchinson2017overcoming}; we take the same view for thermal
conductivity and keep experiment and calculation apart so that the offset can be measured. Our
correction acts on the first-principles calculation itself.

Measurements to correct against are scarce. Of the \num{735} half Heuslers in our corpus with any
thermal-conductivity record, \num{284} have a full first-principles transport calculation and
\num{50} have been measured. Models trained on large public thermal-conductivity datasets are
beginning to appear~\cite{barua2024interpreta}, but \num{50} compounds are too few to learn a
structure--property relation from scratch; they are, as we show, enough to calibrate a correction
to calculated values.

We therefore predict the measured $\kappa_L$ in two steps. A machine-learning model, itself a
surrogate trained on calculations, predicts the full first-principles value from crystal
structure, and a transfer function maps a calculation onto the experimental scale; where a
calculation has been published, the transfer function is applied to it directly. The
\emph{domain of applicability} of the method is cubic half Heuslers with 18 valence electrons
(Section~\ref{sec:domain}). Each measured compound is scored against one reference laboratory
chosen by a fixed rule that never reads a calculation. With a half Heusler written as XYZ
(Section~\ref{sec:method}), a \emph{bonding family} is the set of compounds that share Y and Z,
and a \emph{chemistry cluster} the set that share X. The \emph{blind test} runs the two steps end
to end with each compound's chemistry cluster held out of both; Table~\ref{tab:notation} defines
the terms used throughout.

The paper makes four contributions, taken in the order of the Results
(Section~\ref{sec:results}).
\begin{enumerate}[label=(\roman*), itemsep=1pt, topsep=3pt, leftmargin=2.2em]
  \item \emph{The offset, measured.} Published transport calculations run high in \num{29} of
        \num{34} half Heuslers, and the offset shrinks as temperature rises
        (Section~\ref{sec:offset}).
  \item \emph{A closed-form correction.} A transfer function with a magnitude and a temperature
        exponent (Eq.~\eqref{eq:transfer}), fitted on published calculations, has magnitude
        $c = \num{0.42}$, better constrained than the exponent $p = \num{0.70}$. It carries over
        to a separately computed set of calculations, and a magnitude refitted per bonding family
        is validated held out in five of the nine families with two or more measured members
        (Section~\ref{sec:transferresults}).
  \item \emph{An end-to-end blind test against experiment.} With a model that reproduces published
        calculations to \SI{21.9}{\percent} (reference seed; Section~\ref{sec:mlrepro}), the two
        steps together predict measured $\kappa_L$ in the blind test to a median error of
        \SI{33.1}{\percent} inside the domain of applicability (\SI{37.4}{\percent} over all
        \num{50} measured compounds). Their typical disagreement with measurement is similar in
        size to the spread between two laboratories measuring the same compound. They are closer
        to measurement than null baselines fitted to the individual measurements, which use no
        information about the compound, but not significantly closer than a constant fitted
        compound by compound, as their own constants are (Section~\ref{sec:blind}).
  \item \emph{Predictions that can be tested.} Ten issued predictions with intervals, \num{15}
        flagged candidates with their reasons, \num{10} estimates conditional on a measurement in
        their family, and values from crystal structure alone for \ce{TmPbAu} and \ce{HfTeOs}, which have neither a measurement nor a published full calculation in our corpus (Sections~\ref{sec:issued}
        to~\ref{sec:structonly}).
\end{enumerate}

Section~\ref{sec:method} sets out the data and the tests behind these results, and
Section~\ref{sec:robust} checks the blind test against other choices and baselines.
Section~\ref{sec:discussion} discusses what the offset is, which route to use and where the method
stops, and Section~\ref{sec:conclusions} ends with what should be measured next.

\section{Methods}
\label{sec:method}

\begin{figure*}[t]
  \centering
  \includegraphics[width=\linewidth]{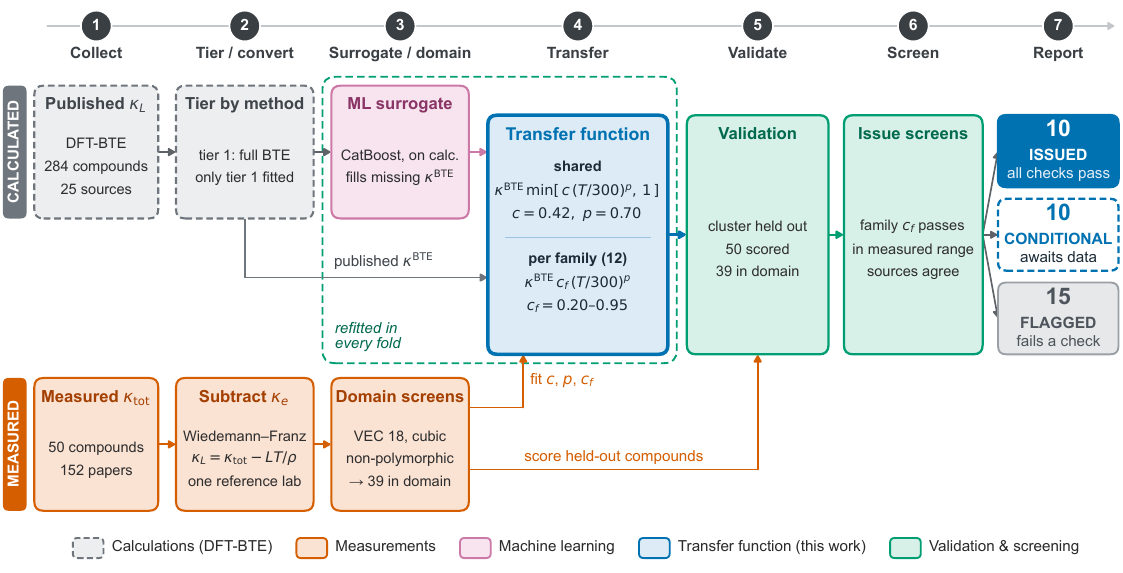}
  \caption{\textbf{Calculations and measurements meet only at the transfer function.} Upper lane
  (grey, dashed): published calculations are collected (step~1) and tiered by method (step~2); only
  tier~1, a full first-principles solution of the Boltzmann transport equation on forces from
  density functional theory (DFT), is
  fitted, and the CatBoost surrogate (pink, step~3), trained on calculations (tier~1 for every half
  Heusler), supplies a calculation only
  where none is published. Lower lane (orange): measured total conductivities from \num{152} papers
  have the electronic part removed by the Wiedemann--Franz law, one reference laboratory per
  compound, and pass the domain screens (\num{39} in domain). The transfer function (blue, step~4) is
  fitted on published calculations against measurements, as shared constants ($c$, $p$) and as one
  magnitude $c_f$ per family (range includes the single-member \ce{Sb}--\ce{Ir} family). Validation
  (step~5) withholds one chemistry cluster from the model and the constants alike (green dashed region,
  refitted in every fold) and scores all \num{50} measured compounds. Candidates (step~6) issue only
  if the family constant is validated or anchored, the value lies in the family's measured range and
  the sources agree; step~7 reports the outcome. The step-1 count is of half-Heusler calculation sources; over all
  Heuslers there are \num{29}.}
  \label{fig:method}
\end{figure*}

\begin{figure*}[t]
  \centering
  \includegraphics[width=\linewidth]{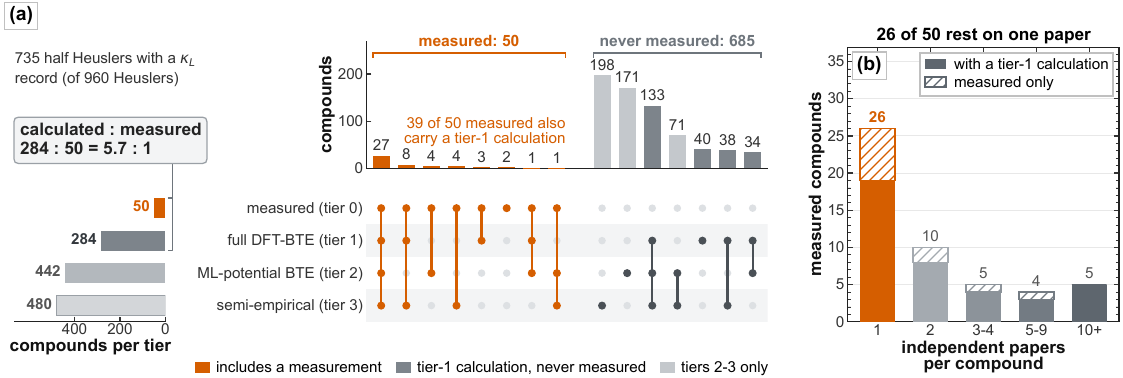}
  \caption{\textbf{The half-Heusler $\kappa_L$ corpus.} (a)~The \num{735} half Heuslers with a
  thermal-conductivity record (of \num{960} Heusler compounds) by evidence type: rows are method tiers
  (measured; full DFT-BTE; ML-potential BTE; semi-empirical), columns the tier combinations
  compounds carry; column heights give compound counts and left bars tier sizes. \num{284} compounds
  carry a full calculation and \num{50} a measurement (\num{5.7}:1); \num{39} of the \num{50} also
  carry a tier-1 calculation (not the same set as the \num{39} in-domain compounds of the blind
  test). (b)~Independent publications per measured compound, a preprint and its article counted
  once; hatched, no tier-1 calculation; the single-publication bar is highlighted in orange.
  \num{26} of the \num{50} rest on a single publication.}
  \label{fig:corpus}
\end{figure*}

Measuring the offset fairly, and then correcting it, needs data in which calculations and
measurements are never mixed and tests in which the compound being predicted is withheld. The
prediction itself, of the \emph{measured} lattice thermal conductivity of a half Heusler, is made in
two steps. A machine-learning model predicts the first-principles (Boltzmann-transport-equation, BTE)
conductivity, $\kappa_L^{\mathrm{BTE}}$, from the crystal structure (Section~\ref{sec:model}), and a
transfer function maps a first-principles value onto the experimental scale
(Section~\ref{sec:transfer}). Where a published calculation exists, the transfer function is applied to it directly and the model
is not used. We call the model followed by the transfer function the \emph{model route}, and the
transfer function applied to a published calculation the \emph{published route}. Measurements and calculations are kept apart from the
corpus onward and meet only at the transfer function (Figure~\ref{fig:method}). Rules given here in
outline are stated in full in the Extended Methods (Section~S1 of the Supplementary material).
Section, table and figure numbers that begin with S refer to the Supplementary material throughout.

\paragraph{Notation and conventions} A half Heusler is written XYZ: X is the most electropositive
element (a rare earth or an early transition metal), Y the late transition metal and Z the
main-group element. A bonding family is labelled by its Y and Z elements in no meaningful order, as
in \ce{Sb}--\ce{Pt}. A \emph{member} of a family is a compound of that family with both a usable
reference measurement and a temperature-matched published calculation. Formulas list the elements in order of increasing electronegativity, as the
descriptors do, so the compound often written TmPtSb in the literature appears here as \ce{TmSbPt}.
The \emph{adopted} constants are those used for the issued predictions. Table~\ref{tab:notation}
collects the symbols and labels, including the three prediction statuses
(Sections~\ref{sec:screening} and~\ref{sec:refusals}).

\begin{table}[tbp]
  \centering
  \caption{Symbols and terms used throughout.}
  \label{tab:notation}
  \footnotesize
  \setlength{\tabcolsep}{3pt}
  \begin{tabular}{>{\raggedright\arraybackslash}p{0.27\columnwidth}>{\raggedright\arraybackslash}p{0.64\columnwidth}}
    \toprule
    Symbol or label & Meaning \\
    \midrule
    $\kappa_L^{\mathrm{BTE}}$ & lattice thermal conductivity from a full Boltzmann-transport calculation \\
    $\kappa_L^{\mathrm{exp}}$ & a value on the experimental scale (measured, or corrected) \\
    $c$, $p$ & shared magnitude and exponent of the transfer function (Eq.~\eqref{eq:transfer}) \\
    $c_f$ & magnitude fitted to one bonding family (Eq.~\eqref{eq:family}) \\
    $e_i$, $r_i$ & per-compound median error and median prediction-to-reference ratio (Eq.~\eqref{eq:err}) \\
    median error & median over compounds of $e_i$: half of the compounds are predicted more closely \\
    within $2\times$ & share of compounds with $\tfrac12 \le r_i \le 2$ \\
    held out & nothing fitted to the prediction saw the compound's data (or, for a cluster, any of its compounds' data) \\
    seeds; reference seed & the model is trained five times; results are the median over the five unless seed~0, the reference seed, is named \\
    null baseline & a prediction without compound-specific information (a constant, or a power law in $T$) \\
    conformal interval & a range whose width is set by the errors of held-out compounds (Eq.~\eqref{eq:conformal}) \\
    tier 0--3 & measured; full first-principles BTE; approximate BTE; semi-empirical \\
    chemistry cluster & compounds sharing the X element; the unit held out in the blind test \\
    bonding family & compounds sharing the Y and Z elements \\
    validated / calibrated / anchored & family constant passes the held-out test / fits its own papers only, which is not a test / rests on one member \\
    issued / flagged / conditional & prediction made / declined for a named reason / awaits a family measurement \\
    \bottomrule
  \end{tabular}
\end{table}

\subsection{Corpus construction}
\label{sec:dataset}

A correction from calculation to experiment can be fitted only on compounds that have both, so the
corpus is built to find those pairs and keeps calculations and measurements as separate records. It
holds \num{9495} temperature-resolved values for \num{960} distinct Heusler compounds, \num{735} of
them half Heuslers, and every value records its source and how it was obtained.

Measured values come from the Starrydata database of transport data digitised from the
literature~\cite{katsura2025starrydata} and from direct extraction from primary publications by
multimodal large language models (Google Gemini, \texttt{gemini-2.5-flash} and
\texttt{gemini-2.5-pro}; extraction counts in Section~S1 of the Supplementary material). No value enters the corpus on the model's authority alone: each must pass
the physical and provenance checks below.

Calculated values come from AFLOW~\cite{curtarolo2012aflow} (its Automatic Anharmonic Phonon Library,
AAPL~\cite{plata2017an}, and the Debye--Gr\"{u}neisen estimates of its Automatic Gibbs Library,
AGL~\cite{toher2014highthroug},
kept apart by the tiers of Section~\ref{sec:tiering}), the Materials
Project~\cite{jain2013commentary}, JARVIS~\cite{choudhary2020the}, OQMD~\cite{kirklin2015the} and
published phonon-transport studies. Crystal structures come from the same repositories, the
Crystallography Open Database~\cite{graulis2009crystallog} and OPTIMADE-federated
sources~\cite{andersen2021optimade}.

Experiments measure the total thermal conductivity, so most measured rows obtain the lattice part by
Wiedemann--Franz subtraction,
\begin{equation}
  \kappa_L = \kappa_\mathrm{tot} - \frac{L T}{\rho},
  \label{eq:wf}
\end{equation}
with $T$ the temperature, $\rho$ the electrical resistivity and the Lorenz number from the
single-parabolic-band expression of Kim \emph{et al.}~\cite{kim2015characteri},
$L = [1.5 + \exp(-|S|/116)]\times\SI{e-8}{\watt\ohm\per\kelvin\squared}$ with the Seebeck coefficient
$S$ in \si{\micro\volt\per\kelvin}. Rows without a Seebeck coefficient use the degenerate limit,
\SI{2.44e-8}{\watt\ohm\per\kelvin\squared}, which overstates the electronic part of a semiconductor
and so biases its lattice value low. The resistivity of a dense intermetallic lies between
\SI{e-8}{\ohm\meter} and \SI{e-2}{\ohm\meter}. A row outside that range cannot give a valid
subtraction by Eq.~\eqref{eq:wf}; in practice it is a total conductivity filed under a lattice
label, and it is removed.

No result may be counted twice. A value quoted from another paper, a second digitisation of the same
sample and a data deposit that republishes curves are therefore not counted as separate
measurements, and a preprint is merged with its version of record (rules in Section~S1 of the Supplementary material). Figure~\ref{fig:corpus} summarises the corpus. Where a compound has been measured in more than one
publication (Fig.~\ref{fig:corpus}(b)), one measurement has to stand for it.

\subsection{The measured values and the choice of reference}
\label{sec:measured}

The transfer function maps onto an experimental scale, so each compound needs a single reference
measurement, chosen without sight of any calculation. Every reference is a stoichiometric parent
compound (each coefficient within \num{0.02} of an integer). This excludes the doped and alloyed
specimens that dominate the half-Heusler literature, so the correction measures the offset between a
calculation and a measurement of the same compound. That offset combines error in the calculation
with the effect of the specimen; Section~\ref{sec:interpretation} discusses the two.

We never blend the curves of several laboratories, since a blend is a curve nobody measured. A fixed
hierarchy chooses one reference laboratory per compound from the specimen and the temperature
coverage alone: the laboratory closest to the median curve when four or more report the compound,
otherwise the densest specimen, then the widest temperature range. A laboratory that contradicts a
consensus, a data deposit, a nanostructured specimen and a source with fewer than four temperatures
rank last but are never deleted. Within the chosen publication, each \SI{100}{\kelvin} bin takes the
median of its rows.

Two rules adjust this choice. A \emph{family-paper rule} makes one publication that measured two or
more members of a family their common reference, so that a family constant (Section~\ref{sec:transfer}) describes one kind of
specimen. It is dropped in the blind test and for the compound held out in the family test (both described
in Section~\ref{sec:nulls}), so that
a score never depends on which family-mates have been measured. A curve-shape test removes from the
calibration any curve whose slope signals a mis-subtracted electronic term. In the blind test a
compound whose every curve fails keeps one, ranked last, so that the scored population does not
depend on the test.

Of the \num{50} measured compounds, \num{26} have a single publication, so the hierarchy decides the
reference only for the rest. The sources record little about each specimen, so the correction
describes published practice on average, not the microstructure of any one specimen. The spread
between laboratories, against which the scores are read, is given in Section~\ref{sec:blind}.

\subsection{Method tiers and quality control}
\label{sec:tiering}

The calculated values need the same care as the measured ones. The transfer function is fitted to
full first-principles calculations only, so every row carries a
\emph{method tier} that records how its value was obtained. Tier~0 is a laboratory measurement
(\num{50} half Heuslers from \num{152} publications). Tier~1 is a full solution of the phonon
Boltzmann transport equation from first-principles third-order force constants (\num{284} half
Heuslers). Tier~2 is an approximate or reduced transport route, including a transport calculation
whose force constants come from a machine-learned interatomic potential. Tier~3 is a semi-empirical
or quasi-harmonic estimate: Slack~\cite{slack1979the}, Debye--Callaway~\cite{callaway1959model} or
AGL. The tier is assigned from the method string, never from a repository's label.

Five quality-control conditions, given with their reasons in Section~S1, are applied when the corpus
is generated: a cubic $C1_b$ record (space group 216) from a source that is not a
prototype-substitution database; a polymorph flag; separation of half from full Heuslers by the site
ratio; a \SI{200}{\kelvin} temperature floor, below which a boundary-free calculation and a
boundary-limited measurement are not the same quantity; and, for tier~1, a positive match for a
transport solution in the method string with no disclaimer attached. Only tier~1 enters the
calibration and, for half Heuslers, the training of the model.

\subsection{Domain of applicability}
\label{sec:domain}

The tiers settle which values are used; the domain settles which compounds. A transfer function for
\emph{lattice} thermal conductivity is meaningful only where lattice conduction dominates. The domain is therefore half Heuslers with 18 valence electrons, a cubic
$C1_b$ structure, and no other polymorph on record. Transition-metal half Heuslers are
semiconducting at a valence-electron count (VEC) of 18~\cite{graf2011simple}; away from it a large
electronic term must be removed from every measurement. $\mathrm{VEC}=18$ is also where the corpus
has support. In the count, lanthanides and actinides are taken as group~3 (full rule in Section~S1 of the Supplementary material). Rare-earth members based on Pt or Pd at $\mathrm{VEC}=18$ are often
narrow-gap or semimetallic, so for them the electronic subtraction is less secure. The structural
conditions follow because the calculation being calibrated is for the cubic phase.

Two further screens apply to prediction candidates only. A candidate containing \ce{Ce}, \ce{Sm},
\ce{Eu} or \ce{Yb} is flagged, not issued. In these elements the 4f level lies close to the Fermi
energy, so the valence can vary from compound to compound~\cite{lawrence1981valence} and differ from
the trivalent value assumed when counting electrons; a compound counted as $\mathrm{VEC}=18$ may
then not be an 18-electron semiconductor. The validation compound \ce{YbNiSb} is still scored. A
candidate containing a radioactive element is excluded, since the corpus holds no training example
of one.

All \num{50} measured half Heuslers are scored in the blind test; \num{44} have $\mathrm{VEC}=18$
and \num{39} also pass the structural conditions. These \num{39} form the in-domain set, and the
same screens are applied to every prediction candidate. Table~\ref{tab:populations} lists the
populations that recur in the paper; results for different populations are not directly comparable.

\begin{table}[tbp]
  \centering
  \caption{Populations at a glance. Every compound is a half Heusler; each row is a subset of the
  row above it, except where noted.}
  \label{tab:populations}
  \footnotesize
  \setlength{\tabcolsep}{3pt}
  \begin{tabular}{r>{\raggedright\arraybackslash}p{0.47\columnwidth}>{\raggedright\arraybackslash}p{0.32\columnwidth}}
    \toprule
    $n$ & Population & Used for \\
    \midrule
    \num{735} & any $\kappa_L$ record & corpus (Fig.~\ref{fig:corpus}) \\
    \num{284} & full transport calculation (tier~1) & model training labels \\
    \num{230} & tier~1 with $\mathrm{VEC}=18$ & model against calculation \\
    \num{50}  & measured (tier~0)$^{a}$ & blind test, unscreened \\
    \num{39}  & measured, inside the domain & blind test, main result \\
    \num{34}  & in domain, with a matched published calculation & offset; shared constant \\
    \num{31}  & of the \num{34}, with a measured family-mate & family test \\
    \bottomrule
    \multicolumn{3}{p{0.92\columnwidth}}{\footnotesize $^{a}$~A subset of the \num{735}, not of the
    \num{284}: \num{11} of the \num{50} have no full calculation.} \\
  \end{tabular}
\end{table}

\subsection{Training data, descriptors and the machine-learning model}
\label{sec:trainingdata}
\label{sec:model}

The first of the two steps is the model. Where no calculation is published it must supply one, so it
is trained on calculations and never on measurements. Its label is the phonon Boltzmann-transport conductivity of a cubic crystal,
\begin{equation}
  \kappa_L^{\mathrm{BTE}} \;=\; \frac{1}{3N_{\mathbf{q}}V}\sum_{\mathbf{q},s}
    C_{\mathbf{q}s}\, v_{\mathbf{q}s}^{2}\, \tau_{\mathbf{q}s},
  \label{eq:bte}
\end{equation}
a sum over the $N_{\mathbf{q}}$ sampled wavevectors $\mathbf{q}$ and branches $s$ of the mode heat
capacity $C_{\mathbf{q}s}$, squared group velocity $v_{\mathbf{q}s}^{2}$ and lifetime
$\tau_{\mathbf{q}s}$, divided by the cell volume $V$. The training pool holds \num{681} labels on
\num{463} compounds; of the labels, \num{544} are at \SI{300}{\kelvin} and \num{119} at
\SI{500}{\kelvin}. Of the compounds, \num{284} are half Heuslers (\num{230} with
$\mathrm{VEC}=18$), each labelled by a tier-1 calculation. Full Heuslers share the descriptors and
also enter the pool; those without a transport calculation enter through a semi-empirical estimate
at reduced weight. Held-out scores are always computed on half Heuslers alone. The labels are
strongly right-skewed, so the model is fitted on $\log_{10}\kappa_L^{\mathrm{BTE}}$ (distribution in
the Extended Methods).

The model is a set of gradient-boosted decision trees (CatBoost~\cite{prokhorenkova2017catboost}, chosen among four regressors in
Section~\ref{sec:modelselection}). Each prediction is the median over five random seeds. A few
diagnostics use seed~0 alone, the \emph{reference seed}, and are labelled as such; all main results
are medians over the five seeds.

\paragraph{Descriptors} Each row is described by \num{45} descriptors of the composition, the cubic
cell and the temperature (grouped in Section~S1 of the Supplementary material): the mass, radius and
electronegativity of X, Y and Z, and combinations of these across sites; the cell geometry, with a
full-Heusler flag; bond lengths, left missing where they cannot be computed; and temperature, as
$T$, $1000/T$ and $\log_{10}T$. The lattice parameter itself is never a descriptor, because sources
mix the primitive and conventional cells.

\subsection{Model selection}
\label{sec:modelselection}

To see how much the choice of regressor matters, we compared CatBoost, LightGBM, a Gaussian process
and kernel ridge regression on identical rows and descriptors. The two tree models also receive the
same row weights; the two kernel models are fitted unweighted. Each was scored out of fold
with five-fold cross-validation grouped by element system (the set of elements in a compound). CatBoost has the highest $R^2$ and the lowest root-mean-square error (RMSE) and mean absolute error (MAE) (Table~\ref{tab:models}; the scores are
defined in Section~\ref{sec:metrics}).
LightGBM has the lowest median error per compound, but by only \num{0.1} percentage points, too
little to separate the two, and we retain CatBoost. The Gaussian process is close behind and kernel
ridge clearly worse, so the choice among the leading three matters little for the fit to
calculations. These scores measure fit to calculated labels over the whole pool, on element-system
folds and not the X-site clusters of the validation, so they are not comparable with the
half-Heusler accuracy of Section~\ref{sec:mlrepro}. Parity plots for all four are in Section~S4 of
the Supplementary material.

\begin{table}[tbp]
  \centering
  \caption{Model selection: four regressors trained on identical rows (\num{681} rows, \num{463}
  compounds) and scored out of fold on identical element-system folds. RMSE, MAE and $R^2$ are per
  row on $\log_{10}\kappa_L$; the median error is per compound
  (Section~\ref{sec:metrics}). The retained model is in bold.}
  \label{tab:models}
  \small
  \begin{tabular}{lcccc}
    \toprule
    Regressor & RMSE & MAE & $R^2$ & median error \\
    \midrule
    \textbf{CatBoost}  & \textbf{\num{0.270}} & \textbf{\num{0.182}} & \textbf{\num{0.701}} & \SI{26.1}{\percent} \\
    LightGBM           & \num{0.273} & \num{0.186} & \num{0.695} & \SI{26.0}{\percent} \\
    Gaussian process   & \num{0.281} & \num{0.190} & \num{0.676} & \SI{26.4}{\percent} \\
    Kernel ridge       & \num{0.343} & \num{0.236} & \num{0.516} & \SI{32.3}{\percent} \\
    \bottomrule
  \end{tabular}
\end{table}

\subsection{The transfer function}
\label{sec:transfer}

The second step takes a calculation, published or supplied by the model, onto the experimental
scale. A laboratory specimen scatters phonons at grain boundaries, point defects and pores that a
calculation of the perfect crystal lacks; for the \ce{XNiSn} half Heuslers, grain-boundary and
point-defect scattering have been compared directly against measurement~\cite{schrade2017the}. Its
conductivity is therefore lower, and the
specimen contribution should grow as temperature falls and intrinsic mean free paths lengthen.

\paragraph{The shared form} We describe the offset empirically with two parameters:
\begin{equation}
  \kappa_L^{\mathrm{exp}}(T) \;=\; \kappa_L^{\mathrm{BTE}}(T)\;
    \min\!\left[c\left(\frac{T}{300}\right)^{p},\,1\right].
  \label{eq:transfer}
\end{equation}
Here $c$ sets the room-temperature reduction and $p$ its temperature dependence. The cap at one is a
modelling choice for the shared form, which keeps a calibrated value from exceeding its calculation;
it is not a physical requirement, since part of the offset is error in the calculation, which can
have either sign. The constants are fitted on published tier-1 calculations, each paired with the
reference measurement of the same compound, with no model involved. Such pairs exist for
\num{34} in-domain compounds (\num{429} pairs). The fit weights every compound equally:
\begin{equation}
  (c, p) \;=\; \operatorname*{arg\,min}_{c,\,p}\;
    \operatorname*{median}_{i \in \mathcal{C}}\;
    \operatorname*{median}_{j}
    \left|\log_{10}\frac{\hat{\kappa}(T_{ij};c,p)}{\kappa(T_{ij})}\right|,
  \label{eq:fit}
\end{equation}
where $\mathcal{C}$ is the set of compounds used in the fit, $\hat{\kappa}$ the corrected
calculation and $\kappa$ the reference measurement, compared at the temperatures $T_{ij}$ where both
exist. In words, the fit chooses the $c$ and $p$ that bring the corrected calculations closest to
measurement. For each compound we take the absolute logarithm of the ratio of corrected calculation
to measurement at each matched temperature, and the median over those temperatures; we then take the
median over compounds. The logarithm treats a value twice too high and one twice too low as equally
wrong. Taking the median within each compound first makes every compound count once, however many
temperatures it was measured at, and medians keep a single bad point, such as a value misread from a
digitised figure, from pulling the fit.

No compound is scored with a constant that has seen it. Before each compound is scored, the
constants are refitted without it and, where the test requires, without its relatives: without its
whole chemistry cluster in the blind test, without its whole bonding family in the
unseen-family test, and without the compound alone in the family test (the three
tests are described in Section~\ref{sec:nulls}). Fitted on all \num{34} compounds, the constants are $c=\num{0.42}$ and
$p=\num{0.70}$. At \SI{300}{\kelvin} the correction therefore multiplies the calculation by
\num{0.42}; it weakens as temperature rises and disappears at \SI{1036}{\kelvin}, where
$c\,(T/300)^{p}$ reaches one.

The size of the correction is known better than its temperature dependence. Repeating the fit on
\num{200} bootstrap resamples of the compounds, $c$ stays between \num{0.35} and \num{0.55}
(10th--90th percentile), whereas $p$ ranges from \num{0.45} to \num{1.15}. Refitting without one
chemistry cluster or one family at a time gives the same picture (Section~S1). The size of the correction is therefore better determined than its change with temperature;
Section~\ref{sec:limitations} states the consequence: the method gives $\kappa_L$ at a stated
temperature and is not designed to reproduce the shape of $\kappa_L(T)$.

\paragraph{The family form} The offset may differ from one bonding family to another, so every bonding family with at least one
member that has both a published calculation and a usable reference measurement also carries its
own constant:
\begin{equation}
  \kappa_L^{\mathrm{exp}}(T) \;=\; \kappa_L^{\mathrm{BTE}}(T)\;
    c_f\left(\frac{T}{300}\right)^{p}, \qquad p = \num{0.70}.
  \label{eq:family}
\end{equation}
Twelve families qualify. Only the magnitude is refitted, by Eq.~\eqref{eq:fit} restricted to
$c_f<1$ on the family's measured members, because most families are measured over too narrow a
range to identify their own exponent. A family with two or more measured members (nine families) is
tested held out: each member is predicted from the others with its own measurement withheld, and the
shared exponent is refitted without it. One guard acts on the fit and not on the score. In a family
with four or more members (\ce{Sb}--\ce{Pt} and \ce{Ni}--\ce{Sb}), a member whose ratio of calculated
to measured $\kappa_L$ near \SI{300}{\kelvin} lies more than three robust standard deviations from the
family median, on a logarithmic scale and with the spread taken from the median absolute deviation, is
left out when $c_f$ is fitted; it is still predicted and scored. The guard is applied to the family as
a whole, held-out member included, and it acts once, on \ce{YSbPt} in \ce{Sb}--\ce{Pt}
(Section~\ref{sec:familycal}). The family is \emph{validated} only if the median error over
its folds is under \SI{25}{\percent}; otherwise it is \emph{calibrated, not validated}, its constant
describing each member's own reference without transferring between members, and no prediction is
issued through it. A family with one member is \emph{anchored}: its error is an in-sample fit, never
a validation. The family form has no cap. At high temperature it can exceed the calculation by up to
a factor of \num{2.18}, which is where the measurements of those families lie; every issued
prediction carries a factor below one.

A compound receives its family's constant wherever its family has a measured member, and the shared
constant otherwise. The leave-one-chemistry-out blind test is the exception: it is scored with the
shared constant, because a chemistry holdout leaves a compound's family-mates in training and would
not test a family constant at all.

\paragraph{Pairing a calculation with a measurement} Both forms are fitted on pairs of one calculation and one measurement, and the two can be compared
only at the same temperature. Measurements usually cover many temperatures, but most published
calculations are reported at only one. Where a calculation is reported at several temperatures, we
join its points by straight lines on a log--log plot and read off the value we need. Where it is
reported at a single temperature, we extend it by assuming $\kappa_L^{\mathrm{BTE}} \propto 1/T$, the
standard high-temperature behaviour when phonons are scattered mainly by one another (three-phonon
Umklapp scattering). We extend it by no more than a factor of \num{2.5} in temperature either way and
never below \SI{200}{\kelvin}, where this behaviour no longer holds; a value calculated at
\SI{300}{\kelvin} is thus used between \SI{200}{\kelvin} and \SI{750}{\kelvin}. Only one tier-1
calculation in the corpus spans three or more temperatures, so for nearly every pair the temperature
dependence of the calculation is this assumed $1/T$, not a computed one. When a compound has several
independent calculations, we use their median. If they disagree by more than a factor of two at some
temperature, that temperature is \emph{contested} and is not used, for measured compounds and
prediction candidates alike. A compound whose every calculated temperature is contested keeps the one
calculation closest to its family-mates' uncontested values (Section~S1). This is why \ce{TiNiPb} is quoted at \SI{500}{\kelvin}: its two
calculations at \SI{300}{\kelvin} differ by a factor of \num{5.6}.

\subsection{Predictions from crystal structure alone}
\label{sec:structureonly}

For a compound nobody has calculated, the model supplies $\kappa_L^{\mathrm{BTE}}$ and the transfer
function maps it onto the experimental scale. We searched DxMag~\cite{xiao2025highthroug}, AFLOW,
OQMD, JARVIS and the Materials Project for cubic $C1_b$ cells inside the domain and took the lattice
parameter of the lowest-energy site ordering on record. A candidate is predicted only if its family,
its X element and every element are represented among the training half Heuslers. A prediction is
graded \emph{strong} when the model reproduces the family's own published calculations to within
\SI{25}{\percent} held out and the five seeds agree within a factor of \num{1.3}, \emph{moderate}
within \SI{50}{\percent} and \num{1.5}, and \emph{weak} otherwise. Literature status was checked in
OpenAlex~\cite{priem2022openalex}, Crossref and arXiv with a positive control; values reported only
in tables or supplements are outside the reach of that search, and the novelty statements here are
made on that basis. The two steps, structure to calculation and calculation to experiment, carry
errors that we report side by side and never combine. The full procedure and two model-free
cross-checks are in the Extended Methods.

\subsection{Validation, null baselines and prediction intervals}
\label{sec:nulls}
\label{sec:intervals}

Every step above is fitted to data, so each route must be tested on compounds it has not seen. Three
choices define the test: which compounds are hidden from the route, what it is compared against, and
how the uncertainty of each prediction is set.

\paragraph{Protocol} The final evaluation was specified in writing before it was run, and the dated
specification is archived with the code. It fixes how the shared constant is defined and held out,
the rule that each compound has a single reference measurement, how the intervals are built at the
quoted temperature, and which results are reported whatever they show. Earlier development runs on
the same compounds did shape some choices: the rules for building the corpus, the rule for choosing a
family's paper, the uncapped family form and the decision to apply the prediction screens to the
validation set. For that reason the main result is also reported without the screens
(Section~\ref{sec:blind}).

\paragraph{Held-out testing} A model can look accurate simply because it has already seen a close
relative of the compound it is asked to predict. To rule this out, every result for the model route
is obtained with the target compound's \emph{chemistry cluster} withheld. Half Heuslers are grouped by
their X element into chemistry clusters (all Ti half Heuslers together, for example), and a whole
cluster is removed at once, from the model and from the shared constant alike
(\emph{leave-one-chemistry-out}); this is the \emph{blind test}. Full Heuslers in the training set are
grouped by their doubled element and stay in training, so a full Heusler that contains the X element
is not removed. Holding out
whole clusters in this way tests whether a model can extrapolate to chemistry it has not
seen~\cite{durdy2022random}; conventional cross-validation overestimates how well a model does on
new materials~\cite{meredig2018can}. Family-mates with a different X element stay in training, so
the test measures accuracy for a new X element within a known family (numbers of clusters in Section~S1).
In the \emph{family test} the family constants are tested one compound at a time: each compound is
predicted from its family
constant fitted without it and from the shared constant refitted without it, and a sign
test~\cite{dixon1946the} counts how often each comes closer to measurement. In the \emph{unseen-family test} the shared constant is tested on families with no measured member
by removing one whole family at a time from the \num{34} compounds of the published route.

\paragraph{Null baselines} To show that the route has learned something about each compound, we
compare it with two \emph{null baselines} that know nothing about the compound: a single constant
and a simple temperature trend $A(T/300)^{q}$, both fitted to the training measurements. Each
baseline is fitted in two ways. In the first, fixed with the protocol, it is fitted to the individual
measurements: the constant is their median and the trend a least-squares line in logarithms. In the
second it is fitted as the constants of the route are, by minimising the median over compounds of
each compound's median error on a logarithmic scale (Section~\ref{sec:transfer}). The second is the
fairer comparison, because under the first a compound measured at many temperatures weighs more in
the baseline than in the route; it was added after the protocol, and its specification was written
before it was run. A route that cannot beat these has learned nothing
useful. Both baselines use the same held-out chemistry and the same reference measurements as the
route. Each is compared with the route compound by compound with the Wilcoxon signed-rank
test~\cite{wilcoxon1945individual}, which asks whether the route is closer to measurement more often,
and by larger margins, than chance would allow. Because two comparisons are made, the $p$-values are
adjusted with Holm's correction, and the comparison is repeated with each chemistry cluster counted
once, so that a large cluster cannot dominate.

\paragraph{Prediction intervals} Each prediction comes with a \emph{prediction interval}: a range
of conductivity within which the measurement is expected to fall. Its width is set by how far the
route missed on compounds it had not seen. We use the cross-validation (jackknife) form of conformal
prediction~\cite{lei2018distributi}. In plain terms, we collect the held-out errors, sort them by
size, and take the error that a chosen fraction of compounds stay within. Errors are measured as
ratios on a logarithmic scale, so an interval extends by the same factor above and below the
prediction. The errors used are those of the model route, model step included, and the same intervals
are attached to
the issued predictions, which start from a published calculation. Each in-domain compound
contributes one error, in the \SI{100}{\kelvin} temperature bin closest to the quoted temperature,
under the correction it would actually receive: its family constant fitted without it, or else the
shared constant refitted without its chemistry cluster. For $m$ held-out compounds and a target
coverage $\alpha$ (for example $0.9$, meaning nine predictions in ten should fall inside), the
half-width of the interval is the $k$th smallest error,
\begin{equation}
  \delta_\alpha \;=\; \Bigl\{\, \bigl|\log_{10} R_i \bigr| \,\Bigr\}_{(k)},
  \qquad k = \bigl\lceil (m+1)\,\alpha \bigr\rceil,
  \label{eq:conformal}
\end{equation}
where $R_i$ is the ratio of prediction to reference for compound $i$ in that bin. The interval runs
from the prediction divided by $10^{\delta_\alpha}$ to the prediction multiplied by it. The $+1$ in
$k$ is a small-sample correction that makes the interval slightly wider than the plain percentile.

\paragraph{Coverage} Every issued prediction carries such an interval (Table~\ref{tab:predictions}),
and a reader will expect a \SI{90}{\percent} interval to contain the measured value about nine times
in ten. \emph{Coverage}, the fraction of compounds whose measurement actually falls inside their
interval, tests whether it does. It must be measured here, not assumed: split conformal prediction
guarantees its stated coverage under ideal conditions~\cite{lei2018distributi}, but the jackknife
form carries no such guarantee~\cite{barber2021predictive}, and compounds that share a chemistry
cluster or a laboratory are not fully independent. Each compound is therefore tested against an
interval built only from the errors of the other chemistry clusters, and the median over seeds is
used. At \SI{300}{\kelvin} the intervals agree with their nominal levels: the \SI{50}{\percent} interval spans a factor of
\num{1.62} either way and the \SI{90}{\percent} interval a factor of \num{2.85}, and they contain
\num{20} (\SI{51}{\percent}) and \num{36} (\SI{92}{\percent}) of the \num{39} in-domain compounds.
At \SI{500}{\kelvin} they span factors of \num{1.38} and \num{1.89} (all levels in Section~S1).
The test is a weak one: each interval is built from the errors of the same \num{39} compounds with
one cluster removed, so agreement close to the nominal level is expected, and the test would expose
only a cluster whose errors differ markedly from the rest.
Families with one measured member or none are too thin for this test, so their intervals are
untested.

\subsection{Metrics}
\label{sec:metrics}
\label{sec:definitions}

The tests above are scored in two ways. When choosing the machine-learning model, we score how well it
reproduces the calculated values row by row, on the logarithm $y=\log_{10}\kappa_L$, with the usual
RMSE, MAE and $R^2$ (defined in Section~S1). An RMSE of \num{0.270} on the logarithm means the model
typically misses a calculated value by a factor of about $10^{0.270}\approx\num{1.86}$. Every other
number in the paper, including every comparison with measurement, is computed \emph{per compound},
as follows.

The error of a prediction $\hat{\kappa}$ against the reference measurement $\kappa$ at temperature
$T$ is the absolute percentage error,
\begin{equation}
  \mathrm{APE}(T) \;=\; \frac{\bigl|\hat{\kappa}(T) - \kappa(T)\bigr|}{\kappa(T)},
  \label{eq:ape}
\end{equation}
so a prediction of \SI{6}{\watt\per\meter\per\kelvin} against a measurement of
\SI{4}{\watt\per\meter\per\kelvin} has an error of \SI{50}{\percent}. A compound measured at several
temperatures $T_{ij}$ receives a single error $e_i$ and a single ratio $r_i$, the medians over those
temperatures:
\begin{equation}
  e_i \;=\; \operatorname*{median}_{j=1\ldots n_i} \mathrm{APE}(T_{ij}),
  \qquad
  r_i \;=\; \operatorname*{median}_{j} \frac{\hat{\kappa}(T_{ij})}{\kappa(T_{ij})}.
  \label{eq:err}
\end{equation}
From these we report four things. The \emph{median error} is the middle value of $e_i$ over
compounds. The fraction \emph{within a factor of two} counts compounds whose prediction lies between
half and twice the measurement ($\tfrac{1}{2} \le r_i \le 2$). The fractions \emph{within
\SI{30}{\percent}} ($e_i \le 0.30$) and \emph{within \SI{25}{\percent}} ($e_i < 0.25$, the bar a
family must pass to count as validated) count compounds with small errors. The \emph{bias},
$\operatorname{median}_i r_i$, shows whether predictions are too high (above one) or too low (below
one) overall.

Counting each compound once stops a compound measured at many temperatures from outweighing one
measured at few; the correction is fitted the same way for the same reason. The percentage error is
lopsided: predicting twice the measurement counts as \SI{100}{\percent}, predicting half of it as
only \SI{50}{\percent}. The within-a-factor-of-two fraction treats both the same, so it is always
reported beside the median error.

The model is trained five times with different random seeds. The \emph{five-seed median} is the
median of the five per-seed median errors. Where a paired test is quoted ``at the median over
seeds'', the number is the median of the five per-seed $p$-values; it summarises the seeds but is not
itself a $p$-value, so the range over seeds is given where it matters.

\section{Results}
\label{sec:results}

With the data and the tests fixed, the Results first measure the offset between calculation and
measurement (Section~\ref{sec:offset}) and test a two-parameter correction for it
(Section~\ref{sec:transferresults}). They then test the model that supplies a calculation where none
is published (Section~\ref{sec:mlrepro}), the model and the correction together against measurement
(Section~\ref{sec:blind}), and that result against further checks and a stronger baseline
(Section~\ref{sec:robust}). The remaining sections give the predictions the method issues and those
it declines.

\subsection{Published calculations systematically overestimate measured \texorpdfstring{$\kappa_L$}{kappa\_L}}
\label{sec:offset}

\begin{figure*}[t]
  \centering
  \includegraphics[width=\linewidth]{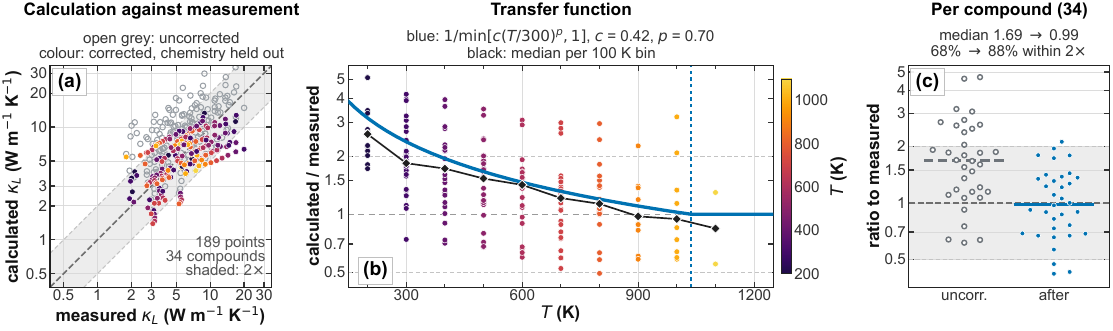}
  \caption{\textbf{The offset between published Boltzmann-transport calculations and measurement,
  and its removal.} (a)~Calculated against measured $\kappa_L$ before (open grey) and after (colour,
  by temperature) the transfer function of Eq.~\eqref{eq:transfer}: \num{189} points, one per
  compound per \SI{100}{\kelvin} bin, from \num{34} in-domain compounds with one reference
  laboratory each; shaded, within a factor of two. (b)~Calculated/measured ratio against $T$, with
  the adopted correction drawn as its reciprocal, $1/\min[c\,(T/300)^p, 1]$ with $c=\num{0.42}$ and
  $p=\num{0.70}$ (blue), and the median of each \SI{100}{\kelvin} bin (black); dotted, the
  \SI{1036}{\kelvin} cap, beyond which \num{3} of the \num{189} points lie. Points below unity,
  where the calculation already lies below the measurement, are out of reach of a downward
  correction. (c)~Per-compound median calculated/measured ratio before and after correction; bars
  mark the medians (\num{1.69} $\to$ \num{0.99}; within $2\times$, \SI{68}{\percent} $\to$
  \SI{88}{\percent}), the statistics quoted in the text. Corrected values in (a) and (c) are held
  out (shared constant refitted without each compound's chemistry cluster); the curve in (b) is the
  adopted constant, fitted on all compounds.}
  \label{fig:offset}
\end{figure*}

A correction is possible only if the offset between calculation and experiment is systematic. It
is. The calculation exceeds the measurement in \num{29} of the \num{34} compounds and in \num{10} of
the \num{12} bonding families, by a median factor of \num{1.69} (each compound's median ratio, then
the median over compounds; Fig.~\ref{fig:offset}(c)). The comparison takes every in-domain
half Heusler that has both a published full calculation and a laboratory measurement, each scored
against its single reference laboratory: \num{34} compounds and \num{429} calculation--measurement
pairs. The
five exceptions, \ce{HfSnPt}, \ce{TiSnPt}, \ce{ZrSnPt}, \ce{ErBiPd} and \ce{LaBiPd}, all belong to
the two families whose median calculation lies below measurement, \ce{Sn}--\ce{Pt} and
\ce{Bi}--\ce{Pd}.

The offset depends on temperature. The median ratio in each \SI{100}{\kelvin} bin falls from
\num{2.61} at \SI{200}{\kelvin} (\num{13} compounds) and \num{1.84} at \SI{300}{\kelvin} (\num{30})
to \num{0.98} at \SI{900}{\kelvin} (\num{11}; Fig.~\ref{fig:offset}(b)), so any correction needs a
temperature term. The bins do not hold the same compounds, and for nearly every pair the calculation
is carried in temperature as $1/T$ (Section~\ref{sec:transfer}), so the trend is that of the
measurements set against an assumed calculated slope. The offset
is also larger than the disagreement between independent calculations of the same compound, a median factor of \num{1.37} over \num{82} half Heuslers (Section~S9 of the Supplementary material), so it is not
the scatter of one code or one group. At a fixed temperature it shows no trend with how conductive
the compound is calculated to be: at \SI{300}{\kelvin} the ratio is uncorrelated with the calculated $\kappa_L$
(Spearman $\rho = \num{0.00}$ over the \num{30} compounds with a pair in the \SI{300}{\kelvin} bin). Section~\ref{sec:interpretation} returns to the
cause. What matters here is that the offset is systematic and can therefore be corrected.

\subsection{A two-parameter transfer function removes the median offset}
\label{sec:transferresults}

We now ask whether the two numbers of Eq.~\eqref{eq:transfer} remove the offset, whether a constant
per bonding family does better, and whether the correction holds beyond the data it was fitted on.

\paragraph{The shared constant} One constant and one exponent, shared by all compounds, remove the
median offset. Fitted on the \num{34} compounds of Section~\ref{sec:offset}, the shared form gives
$c=\num{0.42}$ and $p=\num{0.70}$ (bootstrap ranges in Section~\ref{sec:transfer}). Correcting each
compound with the shared constant refitted without its own chemistry cluster moves the median ratio
from \num{1.69} to \num{0.99} and the fraction of compounds within a factor of two of measurement
from \SI{68}{\percent} to \SI{88}{\percent} (Fig.~\ref{fig:offset}(a) and (c)).

A bias remains at room temperature: values quoted at \SI{300}{\kelvin} through the shared constant
tend to run low. The fit centres each compound across its own temperatures, and the fitted curve lies
above the bin median at every temperature (Fig.~\ref{fig:offset}(b)). At \SI{300}{\kelvin} it
implies \num{2.38} where the bin median is \num{1.84}, which gives the median compound in that bin
\num{0.77} of its measurement. Over all compound--bin
points the corrected median ratio is \num{0.90} (against \num{0.99} per compound, where each compound
is first reduced to its median over temperatures), the same median bias as in the blind test
(Section~\ref{sec:blind}).

\paragraph{One constant per bonding family}
\label{sec:familycal}
\label{sec:deployed}
Does a constant fitted to one bonding family transfer to a member it was not fitted on, and does it
then do better than the shared constant? In five of the nine families with two or more members the
median held-out error is under \SI{25}{\percent}; in the other four it is not. A family member (Section~\ref{sec:method}) is a
compound with both a usable reference measurement and a temperature-matched published calculation.
Each of the \num{12} families with a member carries its own constant
(Eq.~\eqref{eq:family}), tested leave-one-compound-out because an unmeasured target is predicted
the same way, from its measured family-mates.

\begin{figure*}[t]
  \centering
  \includegraphics[width=\linewidth]{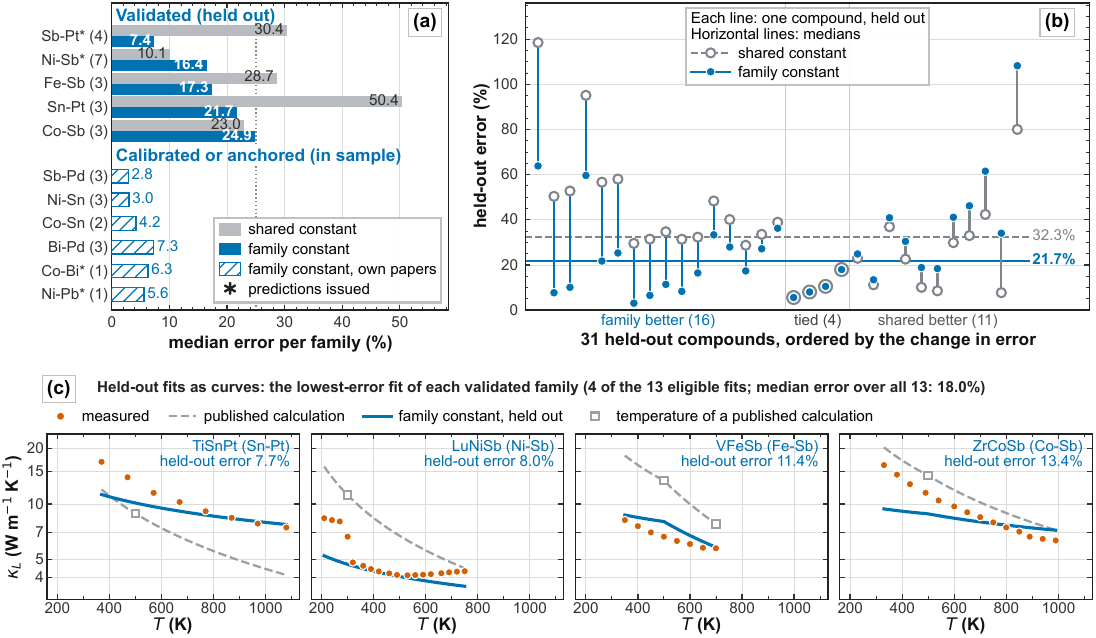}
  \caption{\textbf{Per-family calibration on published calculations.} (a)~Median error per bonding
  family ($n$ in brackets; *, families with issued predictions). Validated families (held-out error
  below \SI{25}{\percent}, the dotted line): each member predicted with its family constant fitted
  without it (blue) and with the shared constant refitted without it (grey). Hatched: the four calibrated families and the two anchored ones (the third single-member family,
  \ce{Sb}--\ce{Ir}, is not drawn; Section~S6), each constant fitted on its family's own reference
  papers and scored on them in sample; held-out figures for the calibrated families are in
  Section~S6 of the Supplementary material. (b)~The same comparison compound by compound, for all
  \num{31} held-out compounds. Each vertical line is one compound and runs from its error with the
  shared constant (open ring) to its error with the family constant (filled dot); the line is blue
  where the family constant is better and grey where it is worse, and a ringed dot marks a tie. The compounds
  are ordered by that change: the family constant is better for the \num{16} on the left, tied
  for \num{4} and worse for the \num{11} on the right (sign test $p=\num{0.44}$). The horizontal
  lines are the two medians. (c)~Held-out family fits drawn as curves, $\kappa_L(T)$. Of the
  \num{13} eligible fits in validated families (at least five temperatures spanning at least
  \SI{300}{\kelvin}), the panel shows the one with the lowest held-out error in each of the four
  families that has an eligible fit. They are the best case of each family and not representative
  ones: the median over all \num{13} is \SI{18.0}{\percent}, and all \num{13} are drawn in
  Section~S5 of the Supplementary material. Points, reference-laboratory measurement; dashed,
  uncorrected published calculation (squares at the calculated temperatures); solid,
  family-calibrated curve with the compound held out.}
  \label{fig:family}
\end{figure*}

Table~\ref{tab:famcal} and Fig.~\ref{fig:family}(a) give the families. The five validated families
have held-out errors from \SI{7.4}{\percent} (\ce{Sb}--\ce{Pt}) to \SI{24.9}{\percent}
(\ce{Co}--\ce{Sb}). The family constant has the lower median error in three of them: it more than halves the error in \ce{Sb}--\ce{Pt} and \ce{Sn}--\ce{Pt}, and lowers it from
\SI{28.7}{\percent} to \SI{17.3}{\percent} in \ce{Fe}--\ce{Sb}.
In \ce{Ni}--\ce{Sb} and \ce{Co}--\ce{Sb} the shared constant does as well or better on the same
compounds; the pre-specified rule still gives every family with measured members its own constant,
so the issued \ce{Ni}--\ce{Sb} values carry \num{0.47} (shared constant: \num{0.42}).

Three cautions apply. First, the data for \ce{Sb}--\ce{Pt} come mainly from one paper, in which
three of the four members are measured at three temperatures each, one fewer than required
elsewhere, and were extracted by the language model, whose only independent check covers four
matched points and is a consistency check, not a validation (Section~S3 of the Supplementary
material). Second, the constant of that family is fitted without \ce{YSbPt}, the one member measured
in another laboratory, which the guard of Section~\ref{sec:transfer} sets aside as an outlier:
predicted from the other three, \ce{YSbPt} has an error of \SI{34.1}{\percent}, larger than with the
shared constant, and the family's \SI{7.4}{\percent} is the median over all four members. Third,
with three members in most families, a family's status can turn on a single compound;
\ce{Co}--\ce{Sb} lies just under the bar.

\begin{table}[tbp]
  \centering
  \caption{Per-family transfer functions, $\kappa_L^{\mathrm{BTE}}\,c_f\,(T/300)^{0.70}$, with the
  shared exponent. \emph{Held out}: each member predicted from its family-mates alone, with the
  family constant and with the shared constant refitted without it (better in bold); a family is
  \emph{validated} when its held-out error is under \SI{25}{\percent}. \emph{Own paper}: the
  constant fitted on each member's own reference and scored on it, in sample, not a test. Read
  across a row: a validated family is tested on compounds it never saw; the other blocks show only
  how well one constant describes its own papers. Each family's spread of member constants and its
  publications are in Section~S5 of the Supplementary material. The twelfth family,
  \ce{Sb}--\ce{Ir}, has one member that its own constant does not reproduce within the bar even in
  sample; it is listed in Section~S6. Predictions are issued only through validated and anchored
  families. The \ce{Sb}--\ce{Pt} constant is fitted on three of its four members (\ce{YSbPt} is set
  aside by the outlier guard of Section~\ref{sec:transfer}); all four are scored.}
  \label{tab:famcal}
  \footnotesize
  \setlength{\tabcolsep}{3pt}
  \begin{tabular}{lcccccl}
    \toprule
    & & & \multicolumn{2}{c}{held out (\si{\percent})} & own paper & \\
    \cmidrule(lr){4-5}
    Family & members & $c_f$ & family & shared & (\si{\percent}) & status \\
    \midrule
    \multicolumn{7}{l}{\emph{Validated: held out}} \\
    \ce{Sb}--\ce{Pt} & 4 & \num{0.61} & \textbf{7.4} & 30.4 & 2.4 & validated \\
    \ce{Ni}--\ce{Sb} & 7 & \num{0.47} & 16.4 & \textbf{10.1} & 4.5 & validated \\
    \ce{Fe}--\ce{Sb} & 3 & \num{0.36} & \textbf{17.3} & 28.7 & 13.9 & validated \\
    \ce{Sn}--\ce{Pt} & 3 & \num{0.89} & \textbf{21.7} & 50.4 & 13.1 & validated \\
    \ce{Co}--\ce{Sb} & 3 & \num{0.42} & 24.9 & \textbf{23.0} & 8.2 & validated \\
    \midrule
    \multicolumn{7}{l}{\emph{Calibrated on own papers: in sample (held-out results in Section~S6)}} \\
    \ce{Bi}--\ce{Pd} & 3 & \num{0.95} & --- & --- & 7.3 & calibrated \\
    \ce{Ni}--\ce{Sn} & 3 & \num{0.32} & --- & --- & 3.0 & calibrated \\
    \ce{Co}--\ce{Sn} & 2 & \num{0.29} & --- & --- & 4.2 & calibrated \\
    \ce{Sb}--\ce{Pd} & 3 & \num{0.24} & --- & --- & 2.8 & calibrated \\
    \midrule
    \multicolumn{7}{l}{\emph{Anchored: one measured member, in sample}} \\
    \ce{Co}--\ce{Bi} & 1 & \num{0.54} & --- & --- & 6.3 & anchored \\
    \ce{Ni}--\ce{Pb} & 1 & \num{0.51} & --- & --- & 5.6 & anchored \\
    \bottomrule
  \end{tabular}
\end{table}

The other four families, \ce{Bi}--\ce{Pd}, \ce{Ni}--\ce{Sn}, \ce{Co}--\ce{Sn} and \ce{Sb}--\ce{Pd},
are \emph{calibrated, not validated}. Their constant fits each member's own paper
(\SIrange{2.8}{7.3}{\percent} in sample, against \SIrange{2.4}{13.9}{\percent} over all nine
families) but does not transfer from one member to another within the \SI{25}{\percent} bar. We
attribute this to differences between the members as made: a nanostructured sample set against a
dense one, or members measured in different laboratories (Section~\ref{sec:interpretation}). Their
held-out results and the reason in each family are in Section~S6 of the Supplementary material, and
no prediction is issued through them.

\paragraph{Pooled over the held-out compounds} Over all held-out compounds together, the family
constant lowers the median error but is not significantly better compound by compound. Thirty-one of
the \num{34} compounds belong to a family with at least one other measured member and can be
predicted held out. On these \num{31} the family constant gives a median error of
\SI{21.7}{\percent}, with \SI{93.5}{\percent} within a factor of two; the shared constant, refitted
without each held-out compound, gives \SI{32.3}{\percent} and \SI{83.9}{\percent}. The family
constant is the better of the two for \num{16} compounds, worse for \num{11} and tied for \num{4}
(two-sided sign test on the \num{27} untied, $p=\num{0.44}$). Figure~\ref{fig:family}(b) draws the
two errors of every compound and shows why the median falls although the counts are close: where
the family constant helps, the gain is often large; where it does worse, the loss is in most
cases smaller.
Section~S7 of the Supplementary material gives the same comparison over all \num{34} compounds,
which mixes in the three single-member fits, and the result for a capped family form.

Figure~\ref{fig:family}(c) draws held-out family fits as curves, for the compounds measured at
five or more temperatures spanning at least \SI{300}{\kelvin}. It shows the best of each
validated family that has such a compound: \ce{TiSnPt} (\SI{7.7}{\percent}),
\ce{LuNiSb}~\cite{ciesielski2020thermoelec} (\SI{8.0}{\percent}), \ce{VFeSb}
(\SI{11.4}{\percent}) and \ce{ZrCoSb} (\SI{13.4}{\percent}). In \ce{TiSnPt}, above about
\SI{400}{\kelvin} the uncapped family form raises the corrected curve above the calculation,
because the measurements lie there; Section~\ref{sec:interpretation} discusses why a measured
value can exceed the calculation. These are best cases. Over all \num{13} such compounds the
held-out error has a median of \SI{18.0}{\percent} and reaches \SI{41.1}{\percent}
(\ce{HfSnPt}), and Section~S5 of the Supplementary material draws every one of them.

\paragraph{The unseen-family test: the shared constant on families never measured}
\label{sec:unseenfamily}
For a family with no measured member only the shared constant applies, so its error there must be
known. Withholding one family at a time from the fit and predicting its members gives, over the
\num{34} compounds, a median error of \SI{33.0}{\percent} with \SI{79.4}{\percent} within a factor
of two and \SI{32.4}{\percent} within \SI{25}{\percent}; per-family medians range from
\SI{12.8}{\percent} (\ce{Ni}--\ce{Sb}) to \SI{116.8}{\percent} (\ce{Sb}--\ce{Pd}). Every
shared-constant estimate in this paper carries this calibration error.

\paragraph{The correction transfers to a separately computed set of calculations}
\label{sec:sdnnff}

All the calculations used so far come from density functional theory. Elemental-SDNNFF
\cite{rodriguez2023millionsca} lets us test the offset and the shared correction on
calculations made a different way. It
solves the same Boltzmann transport equation, but with interatomic forces predicted by a neural
network, and it predicts the phonon properties of \num{11866} Heusler structures. We used none of its
values to train or fit anything, so it is an independent check. We compare it three ways. First, it
agrees closely with the published first-principles calculations for the \num{18} in-domain compounds
both cover: the median ratio is \num{0.99} and the median difference \SI{18.8}{\percent} at
\SI{300}{\kelvin}. Second, it is a median \num{2.01} times higher than measurement for the \num{21}
compounds it shares with our validation set (median error \SI{101.4}{\percent}). A faithful copy of
the calculation therefore inherits the calculation's offset from experiment
(Section~\ref{sec:offset}). Third, our shared correction lowers the error on the same \num{21}
compounds from \SI{101.4}{\percent} to \SI{36.8}{\percent} (paired Wilcoxon $p=\num{0.0033}$). That
correction was fitted only on first-principles calculations, and refitted here without each
compound's chemistry cluster. It therefore works on calculations it was never fitted to.

Two qualifications apply. First, the agreement with first-principles calculations is not fully
independent: for \num{6} of the \num{18} compounds, the published median includes first-principles
values from the SDNNFF paper itself. Second, the model route of Section~\ref{sec:blind} gives
\SI{44.7}{\percent} on the same \num{21} compounds (reference seed). That is closer to measurement
than raw SDNNFF for \num{16} of the \num{21}, but not as close as corrected SDNNFF. The comparison
favours SDNNFF, which calculates each compound directly, whereas our model predicts each compound
with its whole chemistry withheld. Section~S7 lists the overlap between the two sets of calculations
and the temperature windows used to pair values.

Every test in this subsection started from a calculation, because the correction acts on one. Where
none is published, one has to be supplied first.

\subsection{The model reproduces published calculations}
\label{sec:mlrepro}

\begin{figure*}[t]
  \centering
  \includegraphics[width=\linewidth]{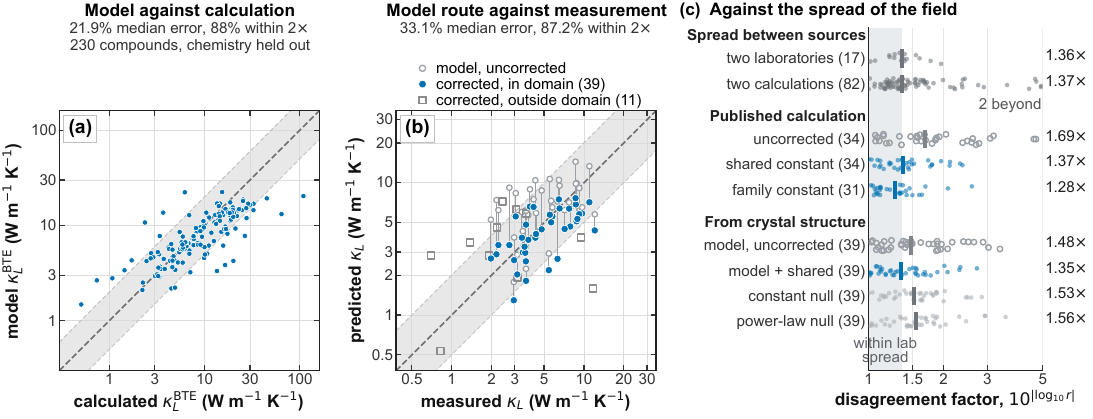}
  \caption{\textbf{From crystal structure to measured $\kappa_L$, and how close that is to the
  spread of the field.} (a)~The model alone, against the published full Boltzmann-transport
  calculation, each compound's chemistry cluster removed from training:
  \SI{21.9}{\percent} median error and \SI{88.3}{\percent} within $2\times$ over \num{230}
  compounds (single seed; per compound over all calculated temperatures); those with a
  \SI{300}{\kelvin} value are drawn. Shaded, within a factor of two. (b)~The model route (model plus shared transfer function)
  against measurement, the chemistry cluster held out of both the model and the transfer function.
  Each of the \num{39} compounds inside the domain of applicability is drawn twice and joined: the
  model's uncorrected value (open grey) and the value after the shared transfer function (blue).
  Open squares, the \num{11} compounds outside the domain, corrected. Each point is the median over
  the compound's matched temperatures. Points are from the reference seed; the printed statistics
  are five-seed medians. (c)~Every route beside the reproducibility of the field: dark grey, the
  spread between two sources; open grey, uncorrected values; blue, values after a transfer function;
  light grey, the null baselines fitted to the individual measurements (fitted per compound they
  reach \num{1.49} and \num{1.54}). Each point is one
  compound's disagreement factor $10^{|\log_{10} r|}$ ($\geq 1$; a factor of two either way plots at
  2); bars mark the median, printed on the right, and the number of compounds is in brackets; two of the \num{82} factors
  between calculations lie beyond the axis.
  \emph{Spread between sources}: median pairwise factor between laboratories measuring an in-domain
  compound, and the factor between independent full calculations at \SI{300}{\kelvin}.
  \emph{Published calculation}: the \num{34} compounds of Fig.~\ref{fig:offset}, uncorrected and
  with the shared constant refitted without each compound's chemistry cluster, and the \num{31}
  scored held out with their family constant. \emph{From crystal structure}: the \num{39} in-domain
  compounds of the blind test (model rows are each compound's median over five seeds) and the two
  null baselines; because each model compound is first reduced to its median over seeds, these
  medians can differ slightly from the symmetric errors of Table~\ref{tab:accuracy}, which take the
  median over seeds of each seed's median. Shaded, below the laboratory median.}
  \label{fig:modelblind}
\end{figure*}

A model can supply the calculation where none is published. Tested on the compounds that have one,
it reproduces them to a median error of \SI{21.9}{\percent} (reference seed). The score is against
the published full calculations of the \num{230} $\mathrm{VEC}=18$ half Heuslers, each predicted
with its chemistry cluster withheld from training (\num{27} clusters); \SI{88.3}{\percent}
lie within a factor of two, and the median ratio is \num{0.87} (Fig.~\ref{fig:modelblind}(a)). The
result covers only the step from structure to calculation. The blind test below already
includes this step, so the two errors should not be compounded.

\paragraph{What the model has learned}
\label{sec:shap}

To see which descriptors the model relies on, we decompose the reference-seed model, fitted on all
\num{681} training rows, with \textsc{shap} (SHapley Additive exPlanations)~\cite{lundberg2017a};
Section~S4 of the Supplementary material gives the attribution in full. The full-Heusler flag
contributes most, followed by size and bond length: a larger X-site radius and longer bonds lower
the prediction. The pattern is consistent with phonon physics, which does not show that the model
has learned that physics. Two limits matter most: the attribution explains the model of the
\emph{calculation}, not of a measurement, which also depends on the specimen; and correlated
descriptors share attribution, so the ranking describes groups of descriptors more reliably than
individual ones.

The correction and the model have now each been tested alone, the correction on published
calculations and the model against them. The next test applies both and compares the result with measurement.

\subsection{From crystal structure to measurement: the held-out test against experiment}
\label{sec:blind}

The central test of the paper asks how close the model route, model plus shared transfer function,
comes to measurement for a compound whose chemistry was withheld from every fitted step. We score
the \num{39} in-domain half Heuslers, each against its single reference laboratory with its chemistry cluster removed from both the model's training and the fit of the shared constant.

The model route has a median error of \SI{33.1}{\percent} (bootstrap 95\,\% interval
\SIrange{23.6}{38.9}{\percent}) and places \SI{87.2}{\percent} of compounds within a factor of two
of measurement, both medians over five model seeds (Table~\ref{tab:accuracy};
Fig.~\ref{fig:modelblind}(b) and (c)). Across seeds the median error runs from \SI{24.6}{\percent}
to \SI{35.4}{\percent} and the within-$2\times$ fraction from \SI{87.2}{\percent} to
\SI{89.7}{\percent}; the reference seed, at which the per-compound analyses below are quoted, gives
\SI{31.7}{\percent}, slightly better than the seed median. Without the domain screens, over all
\num{50} measured half Heuslers, the same protocol gives \SI{37.4}{\percent} (seed range
\SIrange{33.6}{38.9}{\percent}), against \SI{48.2}{\percent} for the uncorrected model (medians over
seeds). The correction moves the median ratio of predicted to measured conductivity from \num{1.40}
to \num{0.90}. Being a multiplier, it cannot remove the compression visible in
Fig.~\ref{fig:modelblind}(b): the lowest-$\kappa_L$ compounds tend to be over-predicted and the
highest under-predicted.

\begin{table*}[t]
  \centering
  \caption{Accuracy of each route against measurement. Errors are per compound; the median error
  carries a bootstrap 95\,\% interval over compounds (\num{5000} resamples). Bold marks the main
  result. Each null baseline is given under both ways of fitting it (Section~\ref{sec:nulls}). The symmetric error
  $|\log_{10} r|$ weights over- and under-prediction equally. Model rows are medians over five seeds
  (seed ranges of the two corrected rows in the text); the nulls and the published-calculation rows do not depend on the seed.}
  \label{tab:accuracy}
  \small
  \begin{adjustbox}{max width=\linewidth}
  \begin{tabular}{llrccc}
    \toprule
    Route & Held out & $n$ & Median error (\si{\percent}) & Within $2\times$ (\si{\percent}) & Median $|\log_{10} r|$ \\
    \midrule
    Model + shared constant, in domain & chemistry cluster & 39 & \textbf{33.1} [23.6, 38.9] & \textbf{87.2} & \textbf{0.132} \\
    Model + shared constant, all measured & chemistry cluster & 50 & 37.4 [30.9, 47.6] & 78.0 & 0.175 \\
    Model, uncorrected, in domain & chemistry cluster & 39 & 41.9 [31.4, 74.5] & 76.9 & 0.182 \\
    Null: constant $\kappa$, fitted to rows, in domain & chemistry cluster & 39 & 46.7 [38.3, 52.8] & 84.6 & 0.184 \\
    Null: power law, fitted to rows, in domain & chemistry cluster & 39 & 45.6 [35.9, 55.7] & 84.6 & 0.192 \\
    Null: constant $\kappa$, fitted per compound, in domain & chemistry cluster & 39 & 35.9 [26.8, 49.8] & 76.9 & 0.172 \\
    Null: power law, fitted per compound, in domain & chemistry cluster & 39 & 42.2 [30.3, 55.7] & 74.4 & 0.188 \\
    \midrule
    Published calculation + shared constant & compound & 31 & 32.3 [28.7, 38.9] & 83.9 & 0.136 \\
    Published calculation + family constant & compound & 31 & 21.7 [16.4, 30.4] & 93.5 & 0.106 \\
    Shared constant, family never measured & family & 34 & 33.0 & 79.4 & --- \\
    \bottomrule
  \end{tabular}
  \end{adjustbox}
\end{table*}

The test covers X-site substitution within families that contain measured compounds; for a family
with no measured member, only the separately reported errors of Sections~\ref{sec:unseenfamily}
and~\ref{sec:mlrepro} apply. The result does not depend on \ce{YNiBi}, whose measured curve fails
the curve-shape test: without it the median error is unchanged at \SI{33.1}{\percent}
(Section~S7 of the Supplementary material, with the fraction within \SI{30}{\percent} and the bias
of every route).

\paragraph{Against the null baselines}
\label{sec:vsnulls}
A route is useful only if it beats a prediction that knows nothing about the compound, and here the
answer depends on how the baselines are fitted. Both null baselines are refitted with the same
chemistry held out and scored against the same references. Fitted to the individual measurements,
as first specified, they have median errors of \SI{46.7}{\percent} and \SI{45.6}{\percent}
(Table~\ref{tab:accuracy}), and the route is closer to measurement than each null for
\SI{71.8}{\percent} of compounds, by a median \num{10.3}
percentage points for the constant null (95\,\% interval \numrange{3.8}{22.9}) and \num{11.2} for
the power law (\numrange{3.5}{22.9}). A paired Wilcoxon signed-rank test gives $p=\num{0.004}$
against each null at the median over seeds, and $p=\num{0.007}$ after a Holm correction for the two
comparisons; on the least favourable seed the Holm-adjusted value is \num{0.056}.

Fitted compound by compound, as the constants of the route are (Section~\ref{sec:nulls}), the
baselines come much closer. The constant then has a median error of \SI{35.9}{\percent}, with
\SI{76.9}{\percent} of compounds within a factor of two, against \SI{33.1}{\percent} and
\SI{87.2}{\percent} for the route. The route is the closer of the two for \SI{56.4}{\percent} of
compounds, and the median paired difference of \num{6.1} percentage points has a 95\,\% interval
that runs from \num{18.3} in favour of the route to \num{4.3} against it; the Wilcoxon test gives
$p=\num{0.24}$ at the median over seeds. The power law has a median error of \SI{42.2}{\percent}
(\SI{74.4}{\percent} within a factor of two); the route is closer for \SI{66.7}{\percent} of
compounds, by a median \num{10.3} points (\numrange{0.8}{19.8}), with $p=\num{0.026}$, or
\num{0.052} after the Holm adjustment. On median percentage error the route is therefore not shown
to beat a constant fitted in the same way. It does place more compounds within a factor of two, a
difference we have not tested.

Two stricter tests weaken the evidence for the first fit as well. Compounds of one chemistry cluster
share a model fold and often a laboratory. With each of the \num{18} clusters counted once and the
baselines fitted to the measurements, the Wilcoxon test still gives $p<0.05$ on every seed before
the Holm adjustment (largest \num{0.048}), but the sign test does not, so the cluster-level evidence
is marginal; with the baselines fitted per compound it is absent (largest $p$ of \num{1.00} for the
constant and \num{0.26} for the power law). Over all \num{50} measured compounds the route still
improves on both nulls fitted to the measurements, more weakly (paired Wilcoxon $p=\num{0.035}$ and \num{0.033} at the reference seed, unadjusted; Section~S7). Fitted to the measurements, the nulls place almost as many compounds within a factor of two as the
route (\SI{84.6}{\percent}) and lose on the typical error; fitted per compound, they come close on the
typical error and place fewer within a factor of two. Neither fit shows the route clearly ahead once
each chemistry cluster is counted once.

\paragraph{The errors set against the reproducibility of the field}
The remaining error can be set beside the spread between two sources that report the same compound.
Figure~\ref{fig:modelblind}(c) places both on one multiplicative scale, the factor
$10^{|\log_{10} r|}$ by which two values differ, whichever is the larger. Two laboratories
measuring the same in-domain half Heusler differ by a median factor of \num{1.36} (\num{17} compounds), and two
independent calculations by \num{1.37}. Uncorrected, published calculations sit outside this range
(\num{1.69}), as do the model's uncorrected values (\num{1.48}) and the two null baselines (\num{1.53} and \num{1.56}; \num{1.49} and \num{1.54} when fitted per compound). With the shared constant, the published calculations (\num{1.37}) and
the model route (\num{1.35}) come down to the size of the laboratory spread, and family constants go
slightly below it (\num{1.28}). The comparison is one of size only: the spread rows compare two
sources with each other, the other rows compare a prediction with one chosen reference, and the
scored compound sets differ. Within that limit, the remaining error is of the order of the
disagreement between measurements. We next ask whether further checks and a stronger baseline change this result.

\subsection{Robustness and comparisons}
\label{sec:robust}

\paragraph{Further checks}
\label{sec:failure}
Further checks support the result; Section~S7 gives each in full, including a correction without
the temperature term and the rank correlation between predicted and measured conductivity. The
eleven compounds the screens exclude behave as a separate population, with a median error of
\SI{86.4}{\percent} against \SI{31.7}{\percent} inside the domain at the reference seed (open
squares in Fig.~\ref{fig:modelblind}(b)). Published semi-empirical estimates
(Slack~\cite{slack1979the} and Debye--Callaway~\cite{callaway1959model}) do far worse, at
\SI{91.6}{\percent} on the \num{22} in-domain compounds that have one.

\paragraph{Where the error falls, and a family-informed baseline}
\label{sec:familynull}
The error is uneven across bonding families. Among families with three or more blind-test
compounds, on the model route at the reference seed, the worst are
\ce{Co}--\ce{Sn} (\SI{59.4}{\percent}) and \ce{Ni}--\ce{Sn} (\SI{55.4}{\percent}), three compounds
each, and the best is \ce{Ni}--\ce{Sb} (\SI{13.7}{\percent}, eight compounds). These counts are of
blind-test compounds, which include measured compounds without a published calculation, so they can
exceed the members of Table~\ref{tab:famcal}; the grouping is descriptive, since the unit held out is the chemistry cluster. A stronger baseline therefore fits a power law to
the measured $\kappa(T)$ of the held-out compound's own family, with no calculation. It is defined
for \num{34} of the \num{39} in-domain compounds (the five exceptions are listed in Section~S7; not
the same \num{34} as the published route). There it gives \SI{25.4}{\percent} against
\SI{31.6}{\percent} for the model route at the reference seed, a difference that is not significant
(paired Wilcoxon $p=\num{0.99}$); Section~\ref{sec:routes} draws the consequence for choosing a
route. These checks leave the main result unchanged, with one
qualification: where a family has been measured, a power law through its own measurements does as
well as the model route. We turn to prediction.

\subsection{Ten predictions for unmeasured compounds}
\label{sec:screening}
\label{sec:issued}

This section states which unmeasured half Heuslers can be given a value, and what supports each.
The lowest median held-out error was that of a published calculation corrected with its family constant
(Section~\ref{sec:familycal}), so we apply that route to half Heuslers that have a published
transport calculation but no measurement. A candidate
is issued only if it meets every condition below:
\begin{enumerate}
  \item \textbf{Inside the domain of applicability} (Section~\ref{sec:domain}), with no radioactive
        element and none of \ce{Ce}, \ce{Sm}, \ce{Eu} or \ce{Yb}.
  \item \textbf{A bonding family whose constant has been checked}: \emph{validated} (held-out error
        under \SI{25}{\percent}) or \emph{anchored} on a single member whose in-sample fit is under
        the same bar, an anchored prediction carrying that label (Table~\ref{tab:famcal}).
  \item \textbf{A full transport calculation}, independent calculations agreeing within a factor of
        two at the quoted temperature.
  \item \textbf{Inside, or only marginally outside, the family's measured range.} A value beyond the
        family's measured span by less than a factor of \num{1.33}, the median disagreement between
        laboratories over all pairs of measurements (\num{1.36} per compound), is issued with a
        \emph{marginal} label; a value further outside is flagged.
        An anchored family has a single member and so defines no span; this condition does not apply
        to it.
\end{enumerate}

\begin{figure}[!t]
  \centering
  \includegraphics[width=\columnwidth]{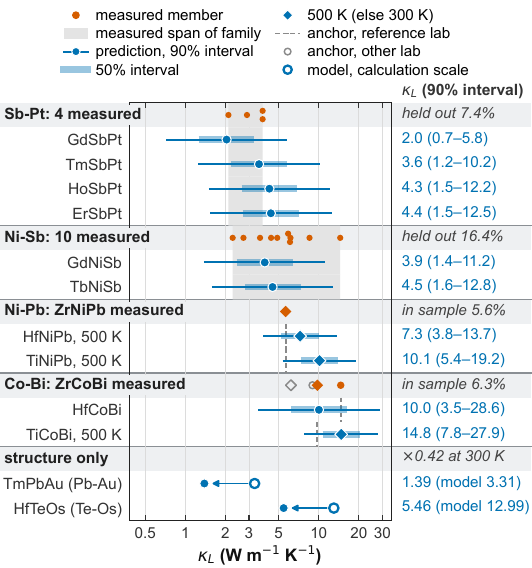}
  \caption{\textbf{Issued predictions beside the measurements that support them.} Each row is one
  issued compound at its quoted temperature (diamonds, \SI{500}{\kelvin}; otherwise
  \SI{300}{\kelvin}), with its \SI{50}{\percent} (thick) and \SI{90}{\percent} (thin) conformal
  intervals. Validated families: orange dots, compounds of the family measured near \SI{300}{\kelvin}; grey shading, the
  span they define. A family's measured count is of every compound of the family measured near
  \SI{300}{\kelvin}; it includes compounds without a published calculation and, for \ce{Ni}--\ce{Sb},
  two outside the domain of applicability (\ce{AlNiSb} and \ce{MgNiSb}), so it can exceed the member
  count of Table~\ref{tab:famcal}. Anchored families (one measured
  member): orange marker and dashed line, that member's reference-laboratory value at the quoted
  temperature (for \ce{Co}--\ce{Bi}, at both \SI{300}{\kelvin} and \SI{500}{\kelvin}); open grey
  marker, a second laboratory's. Bottom: the two structure-only predictions, the model's
  calculation-scale value (open) and the value after the shared transfer function (filled). They
  carry no interval: the model error (Fig.~\ref{fig:modelblind}(a)) and the transfer-function error
  (Section~\ref{sec:unseenfamily}) are reported separately, and interval coverage is not validated for a
  family never measured.}
  \label{fig:predictions}
\end{figure}

Ten compounds meet every condition (Table~\ref{tab:predictions}, Fig.~\ref{fig:predictions}). Three
lie inside the span of a validated family, \ce{Sb}--\ce{Pt} or \ce{Ni}--\ce{Sb}, each differing from
the measured members only on the rare-earth site. Three are marginal extrapolations in
\ce{Sb}--\ce{Pt}; for the six measured compounds lying this close outside their family-mates' span,
the held-out error showed no detectable difference from that of the \num{19} inside it, a comparison
too small to show that the two groups are alike (Section~S7). Four are issued through the
anchored families \ce{Co}--\ce{Bi} and \ce{Ni}--\ce{Pb}, three of them at \SI{500}{\kelvin}, the
temperature of their only uncontested calculation.

\begin{table*}[tbp]
  \centering
  \caption{The ten issued predictions of lattice thermal conductivity for half Heuslers with a
  published transport calculation and no measurement, each through its own family's constant
  (Table~\ref{tab:famcal}). $T$ is \SI{300}{\kelvin}, or \SI{500}{\kelvin} where that is the
  temperature of the only uncontested calculation; $\kappa_L^{\mathrm{BTE}}$ is the median published
  calculation at $T$. The \SI{90}{\percent} intervals are conformal at the quoted temperature; the
  \SI{50}{\percent} intervals, a factor of \num{1.62} (\SI{300}{\kelvin}) or \num{1.38}
  (\SI{500}{\kelvin}) either way, are drawn in Fig.~\ref{fig:predictions}. Their coverage is not
  validated for the anchored families. Span places the value relative to the range of the family's
  compounds measured near \SI{300}{\kelvin}: inside it, or beyond it by the factor shown (marginal).
  An anchored family defines no span.}
  \label{tab:predictions}
  \small
  \begin{adjustbox}{max width=\linewidth}
  \begin{tabular}{llccccc}
    \toprule
    Compound & Family (basis, $c_f$) & $T$ & $\kappa_L^{\mathrm{BTE}}$ & \textbf{Predicted} &
      \SI{90}{\percent} interval & Span \\
    & & (\si{\kelvin}) & \multicolumn{3}{c}{(\si{\watt\per\meter\per\kelvin})} & \\
    \midrule
    \ce{TmSbPt} & \ce{Sb}--\ce{Pt} (validated, \num{0.61}) & 300 & 5.86 & \textbf{3.57} & 1.25--10.17 & inside \\
    \ce{GdNiSb} & \ce{Ni}--\ce{Sb} (validated, \num{0.47}) & 300 & 8.38 & \textbf{3.94} & 1.38--11.23 & inside \\
    \ce{TbNiSb} & \ce{Ni}--\ce{Sb} (validated, \num{0.47}) & 300 & 9.59 & \textbf{4.51} & 1.58--12.85 & inside \\
    \ce{ErSbPt} & \ce{Sb}--\ce{Pt} (validated, \num{0.61}) & 300 & 7.18 & \textbf{4.38} & 1.54--12.48 & \num{1.15} above \\
    \ce{HoSbPt} & \ce{Sb}--\ce{Pt} (validated, \num{0.61}) & 300 & 7.00 & \textbf{4.27} & 1.50--12.17 & \num{1.12} above \\
    \ce{GdSbPt} & \ce{Sb}--\ce{Pt} (validated, \num{0.61}) & 300 & 3.34 & \textbf{2.04} & 0.72--5.81 & \num{1.03} below \\
    \midrule
    \ce{HfCoBi} & \ce{Co}--\ce{Bi} (anchored, \num{0.54}) & 300 & 18.60 & \textbf{10.04} & 3.52--28.61 & --- \\
    \ce{TiCoBi} & \ce{Co}--\ce{Bi} (anchored, \num{0.54}) & 500 & 19.10 & \textbf{14.75} & 7.80--27.88 & --- \\
    \ce{HfNiPb} & \ce{Ni}--\ce{Pb} (anchored, \num{0.51}) & 500 & 9.96 & \textbf{7.26} & 3.84--13.72 & --- \\
    \ce{TiNiPb} & \ce{Ni}--\ce{Pb} (anchored, \num{0.51}) & 500 & 13.91 & \textbf{10.14} & 5.37--19.16 & --- \\
    \bottomrule
  \end{tabular}
  \end{adjustbox}
\end{table*}

The intervals may be too narrow where a prediction rests on one calculation. They are the conformal
intervals of Section~\ref{sec:intervals} and vary little across the five seeds (Section~S7). The
calculations being calibrated come from published high-throughput
studies~\cite{li2025highthroug,ohnishi2026database,miyazaki2021machine,carrete2014finding,trans2022lattice},
most from one~\cite{miyazaki2021machine}, and seven of the ten rest on a single calculation at the
quoted temperature. Independent calculations of the same half Heusler disagree by a median factor of
\num{1.37}, and the interval, built from the held-out errors of the model route, does not add
the spread among published calculations, so a single-calculation interval may understate the
uncertainty. Nine of the ten have an uncontested calculation at only one temperature (\ce{HfCoBi}
has two), so no other temperature is quoted.

\ce{GdNiSb} also carries a structural caveat the databases do not record: it has been reported in a
high-temperature \ce{AlB2}-type hexagonal form beside the low-temperature \ce{MgAgAs}-type cubic
one~\cite{skolozdra1997magnetic}, and the prediction is for the cubic phase. Not every candidate
with a corrected value meets the four conditions; we turn to those that do not.

\subsection{Flagged compounds and the measurements that would resolve them}
\label{sec:refusals}

Fifteen compounds carry a corrected value but are declined (\emph{flagged}); Section~S11 of the
Supplementary material tabulates each with the condition it fails and draws every status on one axis.
Most belong to a family calibrated on its papers but not validated held out: nine to
\ce{Bi}--\ce{Pd} and two to \ce{Sb}--\ce{Pd}. \ce{LaSbPt} and \ce{ScSbPt} are clear extrapolations
in a validated family: \ce{LaSbPt} lies a factor of \num{2.00} below the measured \ce{Sb}--\ce{Pt}
floor and \ce{ScSbPt} \num{1.97} above its ceiling. The last two belong to \ce{Sb}--\ce{Ir}, whose
single measured member its own constant does not reproduce within the bar. The range criterion is
deliberately conservative: the two held-out compounds that lay further outside their family's span
were predicted to a median of \SI{18.4}{\percent}, too few cases to support issuing clear
extrapolations.

Ten further estimates, in \ce{Ni}--\ce{Bi}, are conditional, because the family cannot be
calibrated. Its only measured member, \ce{YNiBi}, has a thermal conductivity with a significant
bipolar contribution, as its authors report~\cite{li2015synthesis}, and the lattice term we extract
from it rises with temperature. On the shared constant the ten lie at
\SIrange{2.21}{4.95}{\watt\per\meter\per\kelvin} at \SI{300}{\kelvin}; Section~S12 of the
Supplementary material lists them with the two measurements that would allow nine of them to be
issued. Sections~S13 to~S15 give the status of \ce{YNiBi} and the measurements that would release or
strengthen the other flagged, anchored and structure-only values. One of the ten, \ce{SmNiBi}, also carries a caveat, because samarium's valence can vary
between compounds; it is counted as conditional, not among the fifteen flagged. Across the three
statuses the conformal interval has the same relative width; what changes is the evidence behind the
value, not its stated precision.

For one family a measurement of the cubic phase may not be possible at all. The cobalt arsenides
\ce{XCoAs} (X = Ti, Zr, Hf) have published calculations but no measured family member, and the two
we found reported as synthesised, \ce{ZrCoAs} and \ce{HfCoAs}, crystallise in the orthorhombic
\ce{Co2Si} type~\cite{kleinke1998crystal}. We therefore carry none of the three as a candidate, so
none has a status; the polymorph screen alone would have caught only \ce{HfCoAs} (Section~S11).

Every candidate value so far starts from a published calculation. What can be said for a compound that has
none?

\subsection{Two predictions from crystal structure alone}
\label{sec:structonly}

A value can still be given from crystal structure alone (Section~\ref{sec:structureonly}), but it is
less certain than an issued prediction. Each carries two separately tested errors, stated side by
side and not combined: \SI{21.9}{\percent} for the model step (reference seed;
Section~\ref{sec:mlrepro}) and \SI{33.0}{\percent} for the shared constant on a family never
measured (Section~\ref{sec:unseenfamily}). No conformal interval is given, because its coverage is
not validated for a family never measured. Two compounds are graded strong on the evidence for their
model step; Fig.~\ref{fig:predictions} (bottom) gives both values for each.

For \ce{TmPbAu} (\ce{Pb}--\ce{Au} family; OQMD cubic cell of the lowest-energy ordering,
$a=\SI{6.782}{\angstrom}$), the six compounds of the family with published calculations are reproduced held
out to between \SI{1}{\percent} and \SI{13}{\percent} (median \SI{8.6}{\percent}), and the seeds
agree within a factor of \num{1.11}. The model gives \SI{3.31}{\watt\per\meter\per\kelvin} and the
shared constant ($c=\num{0.42}$) \SI{1.39}{\watt\per\meter\per\kelvin} at \SI{300}{\kelvin}. A family neighbour, \ce{TbPbAu}, has been grown as single
crystals in the cubic half-Heusler structure~\cite{agarwal2026thermodyna}.

For \ce{HfTeOs} (\ce{Te}--\ce{Os} family; AFLOW cubic cell of the lowest-energy ordering,
$a=\SI{6.33}{\angstrom}$), the strong grade rests on one member, \ce{TiTeOs}, reproduced to
\SI{0.5}{\percent} held out; the seeds spread by a factor of \num{1.23}. The model gives
\SI{12.99}{\watt\per\meter\per\kelvin} and the shared constant
\SI{5.46}{\watt\per\meter\per\kelvin}. As a cross-check, the SDNNFF value for its chemical twin
\ce{ZrTeOs} (Zr in place of Hf)~\cite{rodriguez2023millionsca}, carried to hafnium by the twin ratio of the Extended
Methods, gives \SI{14.06}{\watt\per\meter\per\kelvin}; the model value is \num{0.92} times this. The
twin ratio was validated on calculations (median error \SI{12.2}{\percent}), not on a
machine-learned input, so that figure does not carry over to this check.

Section~S16 of the Supplementary material reports the other compounds the search found: further
compounds with no published $\kappa_L$, a group with only a machine-learned value from a published
screen, and two compounds in calibrated families that are flagged. The Discussion
(Section~\ref{sec:discussion}) takes up what these results mean physically, which route to use and how
far the routes reach.

\section{Discussion}
\label{sec:discussion}

The Results show that published calculations exceed measurement by an offset that is systematic
and shrinks as temperature rises, and that a two-parameter transfer function removes the median
offset for compounds held out of its fit. With a model supplying the calculation, the model route
disagrees with measurement by an amount similar in size to the spread between two laboratories
measuring the same compound. None of this says what the offset is. We take that question first,
then ask which route to use for a given compound and how far the routes reach.

\subsection{What the offset and the constants mean}
\label{sec:interpretation}

What causes the offset, and do the fitted constants have a physical meaning? Several effects could
contribute to the offset, and each would leave a different signature (Table~\ref{tab:causes}). The
first is the specimen. A real specimen scatters phonons at defects that the calculation of a
perfect crystal lacks: interstitials and antisites, imaged directly in \ce{TiNiSn}-based
alloys~\cite{kim2017direct}, and grain boundaries~\cite{schrade2017the}. In calculations for
\num{54} half-Heusler alloys, alloy and grain-boundary scattering significantly reduce the
intrinsic conductivity~\cite{berland2019thermoelec}. Such scattering is nearly independent of
temperature, so it matters less as Umklapp scattering strengthens. The second is four-phonon
scattering, which most calculations omit. It lowers the true conductivity, so omitting it biases
three-phonon calculations high~\cite{feng2017fourphonon}; a full treatment of quartic
anharmonicity also shifts phonon frequencies, however, and its net effect differs between
structure types~\cite{xia2020highthroug}. The third is the set-up of the calculation: the
exchange--correlation functional and force-constant cut-offs move a calculation either way. The
fourth lies on the measurement side. An electronic or bipolar term left in $\kappa_L$ by the
Lorenz subtraction raises the measured value~\cite{thesberg2017on,li2015synthesis}, and the
bipolar part grows with temperature as more carriers are excited across the band gap.

\begin{table}[!b]
  \centering
  \caption{Possible causes of the offset and the signature each would leave. ``Ratio'' is
  calculated/measured.}
  \label{tab:causes}
  \footnotesize
  \setlength{\tabcolsep}{3pt}
  \begin{tabular}{>{\raggedright\arraybackslash}p{0.42\columnwidth}>{\raggedright\arraybackslash}p{0.18\columnwidth}>{\raggedright\arraybackslash}p{0.30\columnwidth}}
    \toprule
    Cause & Ratio & As $T$ rises \\
    \midrule
    defects, grain boundaries, pores in the specimen & above 1 & shrinks towards 1 \\
    four-phonon scattering left out of the calculation & usually above 1 & grows \\
    functional, cut-offs, convergence & either & weakly dependent \\
    electronic or bipolar term left in measured $\kappa_L$ & lowers the ratio & effect grows \\
    \bottomrule
  \end{tabular}
\end{table}

Read against Table~\ref{tab:causes}, the three observations of Section~\ref{sec:offset} narrow the
possibilities without settling them. The offset shrinks with temperature, which fits specimen
scattering or a residual electronic term better than missing four-phonon scattering, whose effect
would grow. For scale, if the whole \SI{300}{\kelvin} offset were extrinsic scattering added by
Matthiessen's rule,
$\kappa_L^{\mathrm{exp}}/\kappa_L^{\mathrm{BTE}} = 1/(1+\tau_{\mathrm{int}}/\tau_{\mathrm{ext}})$ in a
grey model, the bin median of \num{1.84} would require an extrinsic scattering rate \num{0.84} times
the intrinsic one. The offset also exceeds the spread between independent calculations, so
differences between codes and settings are not enough to explain it. And at \SI{300}{\kelvin} it
shows no trend with the calculated conductivity: a single boundary mean free path common to every
specimen would suppress the most conductive compounds, whose intrinsic mean free paths are
longest, more strongly. That argues against one common grain size as the whole explanation; defect
populations that differ from specimen to specimen, or errors in the calculations themselves,
remain possible.

The same caution applies to the two constants. Phenomenological temperature dependences
have been fitted to, or tested against, measured half-Heusler curves
before~\cite{geng2014lattice,petersen2015critical}; here we fit instead the \emph{ratio} of
measurement to calculation. The functional form follows that earlier work; what is new here is the
corpus, the class-wide magnitude and its held-out validation. Because nearly every pair carries
the assumed $T^{-1}$ calculated dependence (Section~\ref{sec:transfer}; the one multi-temperature
calculation has a log--log slope of \num{-1.02}), $p$ is in effect the measured slope relative to
it: $p=\num{0.70}$ implies a corrected $\kappa_L \propto T^{-0.30}$.

The correction is empirical, not a boundary-scattering term. In a relaxation-time description of
the conductivity~\cite{callaway1959model}, any temperature-independent extrinsic scattering
(boundaries, point defects or static disorder) added to three-phonon Umklapp scattering caps the
logarithmic slope of the suppression at $1-c$ (\num{0.58} at $c=\num{0.42}$), for a grey or a
spectral model; this bound is our own derivation (Section~S1 of the Supplementary material). The
fitted $p$ exceeds that cap, although its bootstrap range straddles it, and four-phonon scattering
would lower the slope. A measured lattice conductivity above the three-phonon value of the perfect
crystal, as some high-temperature curves show, more likely retains an electronic or bipolar term
that the single-parabolic-band Lorenz number does not remove, or high-temperature measurement
error: the Lorenz number of a multiband or bipolar semiconductor can depart from the single-band
value by a factor of two or more~\cite{thesberg2017on}. Part of the rise of the
measured-to-calculated ratio with temperature, and of the value of $p$, may therefore reflect the
separation of lattice and electronic conduction rather than phonon scattering, and we attach no
mechanism to the parameters.

A family constant describes the specimens behind it, and it may also absorb measurement effects.
The \ce{Bi}--\ce{Pd} constant is close to one, and its rare-earth members are the narrow-gap
compounds for which the electronic subtraction is least secure (Section~\ref{sec:domain}); part of
it may reflect an electronic term left in the measured values rather than better crystals. Several
families rest on a single kind of specimen (the publications behind each family are counted in
Section~S5 of the Supplementary material). All three \ce{Sn}--\ce{Pt} references come from one
paper, as do three of the four in \ce{Sb}--\ce{Pt}, four of the seven in \ce{Ni}--\ce{Sb} and two
of the three in each of \ce{Bi}--\ce{Pd} and \ce{Sb}--\ce{Pd}; there the held-out test partly
measures one paper's internal consistency rather than transfer between laboratories.

The specimens behind the constants also differ in kind. Two \ce{Sb}--\ce{Pd} references
are annealed arc-melted ingots whose authors attribute the near-linear low-temperature
conductivity to additional scattering at grain boundaries, lattice defects and
disorder~\cite{mukhopadhyay2018multifunct}. The two \ce{Co}--\ce{Sn} members were measured on very
different specimens, \ce{NbCoSn} on a near-fully dense pellet~\cite{thiem2025thermoelec} and
\ce{TaCoSn} on a heavily ball-milled nanocrystalline, cobalt-rich specimen that its authors
suggest contains cobalt interstitials~\cite{li2019ntype}; a constant fitted on one does not
transfer to the other (Section~S6). An anchored family inherits its single member's specimen.
\ce{ZrNiPb} was measured only on a ball-milled specimen~\cite{mao2017thermoelec}, so values issued
through \ce{Ni}--\ce{Pb} describe a milled specimen. The \ce{ZrCoBi} reference is a
\SI{98.1}{\percent} dense pellet~\cite{zhao2018synthesis}, whereas the only other laboratory to
measure it reports a value \num{1.7} times lower near \SI{300}{\kelvin} on a far more heavily
milled specimen~\cite{zhu2018discovery}. The \ce{Co}--\ce{Bi} predictions therefore carry a wider
uncertainty than their conformal interval.

\subsection{Which route to use}
\label{sec:routes}

For estimating the measured $\kappa_L$ of an unmeasured half Heusler, the results suggest a simple
order of preference among the routes, set by what is known about the compound. Where a
full transport calculation has been published and the compound's bonding family has been
validated, the family constant gives the lowest median error of the routes tested: over the
\num{31} compounds with a measured family-mate it gives \SI{21.7}{\percent} held out, against
\SI{32.3}{\percent} for the shared constant on the same compounds, a difference that is not
significant compound by compound (Section~\ref{sec:familycal}). Where a calculation exists but
the family has no measured member, only the shared constant applies, with the \SI{33.0}{\percent}
median error measured on families withheld from the fit. Where no calculation exists, the
machine-learning model supplies one. For a new X-site substitution in a family with measured
members, model and shared constant together are tested end to end at \SI{33.1}{\percent} inside the
domain and \SI{37.4}{\percent} over all measured compounds (Section~\ref{sec:blind}); in a family never measured, the value carries two separate errors,
\SI{21.9}{\percent} for the model step (reference seed) and \SI{33.0}{\percent} for the shared
constant.

Where a family has already been measured, a power law fitted to its members' measured $\kappa(T)$
does as well as the model route (\SI{25.4}{\percent} against \SI{31.6}{\percent} for the model
route at the reference seed; Section~\ref{sec:familynull}), but it cannot order compounds within a
family and says nothing where no member is measured. The method therefore adds most where
measurements are sparse.

\subsection{Scope and limitations}
\label{sec:limitations}

The method estimates $\kappa_L$ at a stated temperature. It is not designed to reproduce the shape of
$\kappa_L(T)$: the training labels lie mostly at \SI{300}{\kelvin} and one exponent serves every
compound. Its accuracy is best read against the spread between laboratories (Section~\ref{sec:blind}),
since each compound is scored against one reference laboratory chosen by a fixed rule. Because
compounds share chemistry, the main tests hold out whole chemistry clusters; compounds that also
share a laboratory are discussed in Section~S17. Family constants
apply within families that have measured members, and other families receive the shared constant.
Each issued value inherits the accuracy of the calculation it starts from, and its source is listed so
that it can be updated as new calculations appear. Structure-only values are conditional on the compound forming in the cubic phase. Several alternatives, among them a
two-stage model, a magnitude-dependent calibration term and training on experimental or doped values,
were tested and not adopted because they did not improve held-out accuracy. Section~S17 gives the
quantitative detail behind each of these points.

Several of these limits are set by the measurements available; the Conclusions end with the
measurements that would lift them.

\section{Conclusions}
\label{sec:conclusions}

Published first-principles calculations overestimate the measured lattice thermal conductivity of
half Heuslers by a median factor of \num{1.69}, in \num{10} of \num{12} bonding families, and the
offset shrinks as temperature rises. Defect and boundary scattering added to intrinsic phonon
scattering would give such a trend, and so would an electronic contribution left in the measured
values; the data do not separate them (Section~\ref{sec:interpretation}). Because the offset is systematic,
a two-parameter transfer function corrects it. For compounds with a measured family-mate,
refitting its magnitude within the bonding family lowers the median held-out error from
\SI{32.3}{\percent} to \SI{21.7}{\percent}, although it is the better of the two for only
\num{16} of the \num{27} compounds on which they differ (sign test $p=\num{0.44}$). The
two-parameter correction also carries over to the
calculations of a published machine-learned force field, whose median error against measurement it
lowers from \SI{101.4}{\percent} to \SI{36.8}{\percent}.

Combined with a machine-learning model that predicts the calculated value from crystal structure,
the correction was tested against measurement with each compound's chemistry cluster held out of
every fitted step. Within the domain of applicability the median error is \SI{33.1}{\percent} and
\SI{87.2}{\percent} of \num{39} compounds lie within a factor of two (\SI{37.4}{\percent} over all
\num{50} measured compounds). Inside the domain the typical disagreement with measurement, a
factor of \num{1.35},
is similar in size to the \num{1.36} by which two laboratories measuring the same compound differ,
so further gains are likely to depend on better reference measurements as much as on better
models.

The method turns calculated conductivities into estimates that can be compared directly with
experiment. We issue \num{10} predictions with intervals for unmeasured compounds, and give values
from crystal structure alone for \ce{TmPbAu} and \ce{HfTeOs}, which have neither a measurement
nor a published full calculation in our corpus. Within the domain of Section~\ref{sec:domain} and the limits of
Section~\ref{sec:limitations}, each issued prediction can be tested by a single synthesis and
measurement of the cubic phase.

The measurements that would most strengthen the method are a first measurement in families with no
measured member, such as \ce{Pb}--\ce{Au} and \ce{Te}--\ce{Os}; independent bulk measurements in
families that rest on one compound or one paper; measurements over a wide temperature range on dense
specimens, to constrain the exponent; and two \ce{Ni}--\ce{Bi} measurements free of the bipolar
contribution, which would allow nine of the ten conditional estimates to be issued.

\section*{Data availability}
\label{sec:data-availability}

{\raggedright The following are archived on Zenodo, \repourl:\par}
\begin{itemize}[itemsep=1pt, topsep=3pt, leftmargin=1.5em]
  \item the prediction lists: for every candidate, the calculated and calibrated values, the quoted
        temperature, the prediction intervals, the status (issued, flagged or conditional on a
        measurement) and the reason for that status;
  \item the predictions from crystal structure alone, with the source of each structure;
  \item the fitted transfer-function constants, the per-compound blind-test scores and the output
        of every model seed;
  \item a record of every row that a data-quality rule removed from the corpus or moved to a
        different method tier, each with the rule responsible;
  \item the code that rebuilds the corpus from its sources and generates every figure and number in
        this paper.
\end{itemize}
Candidate structures were taken from public databases (DxMag, AFLOW, OQMD, JARVIS and the Materials
Project). We do not redistribute bulk copies of those databases or of Starrydata. Starrydata is
updated continuously, so a rebuild from its current public release can differ slightly from the
corpus used here; the reference value of every scored compound is archived, and the Supplementary
material names its source.

The measurements analysed here are not ours. Across the whole corpus they come from \num{189}
publications, \num{152} of which lie behind the \num{50} measured half Heuslers. The transport
calculations come from \num{29} publications. The Supplementary material cites every calculation
source and every measurement source but one, so that the provenance of every value can be checked.
They are listed there because they would otherwise outnumber the works discussed in the paper. The
one uncited
measurement source has a digital object identifier (DOI), \texttt{10.36410/jcpr.2020.21.3.319},
that is registered with neither Crossref nor DataCite, so it cannot be resolved into a reference.
Every measured row of the corpus carries the DOI of the work it came from; calculated rows drawn
from public databases carry a source URL instead, because those repositories are not identified by
DOI.

\section*{CRediT authorship contribution statement}

\textbf{Shivam Randive:} Conceptualization, Methodology, Software, Formal analysis,
Data curation, Investigation, Validation, Writing -- original draft.
\textbf{Ojaswini Randive:} Visualization, Writing -- review \& editing.
\textbf{Heet Lungaria:} Visualization, Writing -- review \& editing.
\textbf{Himanshu Pandey:} Supervision, Resources, Project administration,
Writing -- review \& editing.

\section*{Funding}

This research did not receive any specific grant from funding agencies in the public, commercial,
or not-for-profit sectors. Google Cloud computing credits, acknowledged below, covered the cost of
the language-model extraction.

\section*{Declaration of competing interest}

The authors declare that they have no known competing financial interests or personal relationships
that could have appeared to influence the work reported in this paper.

\section*{Acknowledgements}

\acknowledgementtext

\setlength{\bibsep}{0.6pt plus 0.3pt}
\bibliographystyle{elsarticle-num}
\bibliography{references}

\end{document}